\documentclass[aip,jcp,preprint,groupedaddress]{revtex4-1}
\pdfoutput=1
\usepackage[T1,T2A]{fontenc}
\usepackage{amsfonts}
\usepackage{amsmath,amssymb}
\usepackage{graphicx,booktabs}
\usepackage{multirow,dcolumn}
\usepackage{xcolor}
\usepackage[utf8]{inputenc}
\usepackage[english]{babel}

\graphicspath{{./figures/}{C:/Users/Jiri/Dropbox/Papers/Chemistry_papers/2013/Oligothiophenes/Revision_2/revision_2_version_22/complete}}
\newcommand{\RR}{\mbox{\usefont{T2A}{\rmdefault}{m}{n}\CYRYA}}

\begin{document}

\title{On-the-fly ab initio semiclassical dynamics: Identifying degrees of
freedom essential for emission spectra of oligothiophenes}
\author{Marius Wehrle}
\author{Miroslav \v{S}ulc}
\author{Ji\v{r}\'{\i} Van\'{\i}\v{c}ek}
\email{jiri.vanicek@epfl.ch}
\affiliation{Laboratory of Theoretical Physical Chemistry, Institut des Sciences et Ing%
\'{e}nierie Chimiques, Ecole Polytechnique F\'{e}d\'{e}rale de Lausanne
(EPFL), CH-1015, Lausanne, Switzerland}
\date{\today}

\begin{abstract}
Vibrationally resolved spectra provide a stringent test of the accuracy of
theoretical calculations. We combine the thawed Gaussian approximation (TGA)
with an on-the-fly \textit{ab initio} (OTF-AI) scheme to calculate the
vibrationally resolved emission spectra of oligothiophenes with up to five
rings. The efficiency of the OTF-AI-TGA permits treating all vibrational
degrees of freedom on an equal footing even in pentathiophene with $105$%
~vibrational degrees of freedom, thus obviating the need for the global
harmonic approximation, popular for large systems. Besides reproducing
almost perfectly the experimental emission spectra, in order to provide a
deeper insight into the associated physical and chemical processes, we also
develop a novel systematic approach to assess the importance and coupling
between individual vibrational degrees of freedom during the dynamics. This
allows us to explain how the vibrational line shapes of the oligothiophenes
change with increasing number of rings. Furthermore, we observe the
dynamical interplay between the quinoid and aromatic characters of
individual rings in the oligothiophene chain during the dynamics and confirm
that the quinoid character prevails in the center of the chain.
\end{abstract}

\maketitle

%-------------------------------------------------------------------------------

\section{Introduction\label{sec:Introduction}}

%-------------------------------------------------------------------------------
Polythiophenes (T$n$) and their functional derivatives belong among the most
studied compounds among $\pi$-conjugated polymers due to their potential in
organic electronics,\cite%
{Perepichka:2009,*Mishra:2009,*Perepichka:2005,*Klauk:2006} since they
combine remarkable conductivity with excellent thermo- and chemo-stability.
Detailed experimental investigations of polythiophenes have shown that their
optical properties are closely related to the structure of the polymer
backbone: For instance, the $0$--$0$ transition energies are approximately a
linear function of $1/n$, where $n$ is the number of thiophene rings in the
polymer.\cite{Becker:1996,Yang:1997,Yang:1998} Bandgap computations
confirmed validity of this semi-empirical rule for short polymers as well as
its violation for longer chains.\cite%
{Themans:1989,Hutchinson:2003,*Zade:2006}

For a direct comparison with experiments it is, however, crucial to
calculate the vibrationally resolved spectra.\cite%
{Jacquemin:2012,AvilaFerrer:2013} Here, we therefore determine the
vibrationally resolved emission spectra of oligothiophenes T$n$ with two to
five rings, i.e., $n\in{\{2,3,4,5\}}$, since the vibrational line shape is
changing drastically in this range of $n$.\cite{Becker:1996}

The cost of computing a vibrationally resolved spectrum is much higher than
the cost of vertical transition energy calculations since the spectrum
calculation requires the knowledge of the involved potential energy surfaces
(PESs). As it is often difficult to describe PESs accurately in terms of
analytical functions, a popular approach, especially for larger molecules is
to approximate the PESs by harmonic potentials with respect to certain
reference structures.\cite%
{Biczysko:2011,Tatchen:2008,*AvillaFerrer:2012,*Cerezo:2013} The absorption
and emission line shapes of dithiophene have been calculated by Stendardo~%
\textit{et al.\/}\cite{Stenardo:2012} using a double-well potential
describing the torsional mode and global harmonic approximation in the
remaining degrees of freedom. In order to get a good correspondence with
experiment, the authors show that an appropriate choice of these reference
structures is essential, e.g., the ground PES reference structure for the
emission spectrum calculation is found using symmetry constraints.

Alternative strategy employs trajectory-based methods in combination with an
on-the-fly (OTF) \textit{ab initio} (AI) scheme, in which the required
potential energies, forces, and Hessians are computed with an electronic
structure package during the dynamics. It is becoming increasingly clear
that \textit{ab initio} semiclassical dynamics provides a powerful
spectroscopic tool useful, e.g., for evaluating internal conversion rates%
\cite{Ianconescu:2013} or vibrationally resolved spectra.\cite%
{Tatchen_Pollak:2009,Ceotto_AspuruGuzik:2009a,*Ceotto_AspuruGuzik:2009b,*Ceotto:2011a,*Ceotto:2011b,Wong:2011}
Not only do the evolving trajectories provide an intuitive classical-like
picture of the underlying physical and chemical processes, but via
interference, they also partially account for the most important nuclear
quantum effects. The overall computational cost, however, restricts almost
all of these methods to small systems.

As a result, one is forced to strike a balance between physical accuracy and
computational efficiency. In this spirit, OTF-AI Gaussian wave packet
propagation can provide a useful compromise. Within the thawed Gaussian
approximation (TGA), the nuclear wave packet is guided by a central
classical trajectory, which feels the anharmonicity of the potential, while
its width is propagated using the local harmonic approximation.\cite%
{Heller:1975} Hence, the effects of anharmonic or double-well potentials are
partially captured by TGA; moreover, the OTF-AI framework obviates the need
of an a priori knowledge of the landscape of the final PES. More
importantly, due to its moderate computational cost, TGA can treat all
vibrational degrees of freedom on an equal footing even in large systems,
while in smaller systems, it permits using a more accurate electronic
structure description. A well-known shortcoming of the TGA is that it
captures accurately only the short-time dynamics and, therefore, only
describes the broad spectral features.\cite{Heller:1981} Nevertheless, due
to interaction with solvent and other phenomena contributing to spectral
broadening, the experimental spectra are also typically not fully
vibrationally resolved.

Although rewarding, a mere reproduction of an experimental spectrum, no
matter how accurate, does not provide a deeper insight into the associated
physical and chemical processes; it is a careful analysis of the simulation
that can provide such information. The extraction of this essential
information, which is often omitted, can be as difficult as the simulation
itself, especially for larger molecules. For example, explanation of changes
in the vibrational line shape of the spectra due to increasing polymer chain
length, which is done here for oligothiophenes, increases drastically the
complexity of the analysis. Therefore, in addition to providing an efficient
computational protocol for computing vibrationally resolved electronic
spectra we also present a systematic approach for extracting the essential
information about the underlying dynamics. 
%-------------------------------------------------------------------------------

\section{\label{sec:Theory}Theory}

%-------------------------------------------------------------------------------

\subsection{\label{subsec:spec_calc}Emission spectrum calculation}

%-------------------------------------------------------------------------------
In the time-dependent approach pioneered by Heller,\cite{Heller:1981} the
molecular spectrum is determined by the Fourier transform of an appropriate
correlation function. Within the electric dipole approximation,
time-dependent perturbation theory, and rotating wave approximation, the
correlation function required for computing the emission spectrum is 
\begin{equation}
C_{\text{em}}(t)\propto \text{Tr}\,\left[ \hat{\rho}_{1}(T){\hat{\mu}}_{10}%
\hat{U}_{0}(-t){\hat{\mu}}_{01}\hat{U}_{1}(t)\right] .  \label{eq:C_em}
\end{equation}%
Here, $\hat{\rho}_{1}(T)$ is the nuclear density operator in the first
excited electronic state ($S_{1}$) at temperature $T$, $\hat{U}_{j}(t)=\text{%
exp}\left( -i\hat{H}_{j}t/\hbar \right) $ for $j\in \{0,1\}$ denotes the
nuclear quantum evolution operator on the $j$th electronic surface $S_{j}$,
and ${\hat{\mu}}_{ij}$ is the transition dipole moment operator coupling
states $S_{i}$ and $S_{j}$. Within the Franck-Condon approximation and in
the low temperature limit, the correlation function (\ref{eq:C_em}) becomes 
\begin{align}
C_{\text{em}}(t)& \propto \langle \Psi _{\text{init}}|\hat{U}_{0}(-t)\hat{U}%
_{1}(t)|\Psi _{\text{init}}\rangle   \label{eq:C_em_simplified} \\
& =\langle \Psi _{\text{init}}|\hat{U}_{0}(-t)|\Psi _{\text{init}}\rangle
e^{-iE_{1}t/\hbar },  \notag
\end{align}%
where $E_{1}$ is the energy of the ground vibrational state of $S_{1}$.
Equation~(\ref{eq:C_em_simplified}) states\cite{Heller:1981} that
propagation of the ground vibrational state of $S_{1}$ on $S_{0}$ determines
the correlation function $C_{\text{em}}(t)$, and hence the spectrum, which
is obtained via a Fourier transform 
\begin{equation}
\sigma (\omega )=A\omega ^{k}\int C_{\text{em}}(t)e^{i\omega t}dt,
\label{eq:sigma_em}
\end{equation}%
where $k=0$ for the line shape and $k=3$ for the emission spectrum.
Prefactor $A$ is a constant factor depending on the type of spectra.\cite%
{Lami:2011,Biczysko:2011,AvilaFerrer:2013} Since it is constant, in our
calculations $A$ was chosen so that the spectra are normalized in the $%
L^{\infty }$ norm, i.e., the highest spectral peak is of unit intensity. 
%-------------------------------------------------------------------------------

\subsection{\label{subsec:TGA}Thawed Gaussian Approximation}

%-------------------------------------------------------------------------------
The celebrated \textit{thawed Gaussian approximation}\cite%
{Heller:1975,Lee_Heller:1982} of Heller belongs among the earliest practical
semiclassical approaches to quantum dynamics. The main idea is exceedingly
simple---since a Gaussian wave packet (GWP) evolving in a globally harmonic
potential retains its functional form, one expects that propagating a single
thawed GWP using a local harmonic approximation for the potential can
provide a reasonable approximation in many applications, especially when the
dynamics of interst is ultrafast. Although the accuracy of the single GWP
description is clearly limited, it can provide the most important
information beyond that contained in static calculations employing globally
harmonic approximation for the potential.\cite%
{Grossmann:2006,Baranger_Schellhaass:2001}

Within TGA, the evolving GWP is assumed in the form 
\begin{align}  \label{eq:TGA_wavefunction_Ansatz}
\psi^{t}(q)=N^{0}\exp\bigl\{&-(q-q^{t})^{\mathsf{T}}\cdot A^{t}\cdot(q-q^{t})
\notag \\
&+\frac{i}{\hbar}\bigl[p^{t}\cdot(q-q^{t})+\gamma^{t}\bigr]\bigr\},
\end{align}
where $N^{0}$ is a normalization constant, $x^{t}=(q^{t},\, p^{t})$ denotes
the GWP's phase-space center, $A^{t}$ is a real, symmetric width matrix, and 
$\gamma^{t}$ represents an overall phase factor. Note that $\gamma^{t}$ is a
time-dependent complex number the imaginary part of which guarantees
normalization of $\psi^{t}(q)$ for $t\geq0$. The key ingredient of the
method consists in expressing the potential $V(q)$ in the \textit{local
harmonic approximation} (LHA). This in turn yields a time-dependent
effective potential 
\begin{align}  \label{eq:TGA_effective_potential}
V_{\mathrm{eff}}^{t}(q)&=V(q^{t})+\nabla V(q^{t})^{\mathsf{T}}\cdot(q-q^{t})
\notag \\
&+\frac{1}{2}(q-q^{t})^{\mathsf{T}}\cdot\nabla^{2}V(q^{t})\cdot(q-q^{t}),
\end{align}
where the potential $V$, gradient $\nabla V$, and Hessian $\nabla^{2}V$ are
evaluated at the current coordinate center $q^{t}$ of the evolving GWP at
time $t$. As already alluded to above, the second-order Taylor expansion (%
\ref{eq:TGA_effective_potential}) ensures that the ansatz (\ref%
{eq:TGA_wavefunction_Ansatz}) is plausible even for $t>0$. Denoting by 
\begin{equation}
H_{\mathrm{eff}}^{t}\,:=p^{\mathsf{T}}\cdot(G/2)\cdot p+V_{\mathrm{eff}}^{t}
\end{equation}
the effective Hamiltonian and inserting the ansatz (\ref%
{eq:TGA_wavefunction_Ansatz}) into the TDSE 
\begin{equation*}
i\hbar\frac{\partial}{\partial t}\psi^{t}(q)=H_{\mathrm{eff}}^{t}\psi^{t}(q),
\end{equation*}
gives equations of motion for $x^{t}$, $A^{t}$, and $\gamma^{t}$: 
\begin{subequations}
\begin{eqnarray}
\dot{x^{t}} & = & \{x,\, H_{\mathrm{eff}}^{t}\},  \label{eq:TGA_x} \\
\dot{A^{t}} & = & -2i\hbar\, A^{t}\cdot G \cdot A^{t}+\frac{i}{2\hbar}%
\nabla^{2}V(q^{t}),  \label{eq:TGA_mat_A} \\
\dot{\gamma^{t}} & = & \mathcal{L}_{\mathrm{eff}}^{t}-\hbar^{2}\,\text{Tr}%
\bigl[G\cdot A^{t}\bigr],  \label{eq:TGA_equation_of_motions}
\end{eqnarray}
where $G$ is the inverse of the mass matrix and $\mathcal{L}_{\mathrm{eff}%
}^{t}$ denotes Lagrangian dual to $H_{\mathrm{eff}}^{t}$. Numerical
integration of the classical equations of motion (\ref{eq:TGA_x}) is easily
carried out in a symplectic fashion (see Sec.~\ref{sec:comp_details}). In
order to integrate Eq.~(\ref{eq:TGA_mat_A}), we follow the strategy\cite%
{Lee_Heller:1982} proposed by Lee and Heller. Within their method, the
matrix $A^{t}$ is factorized using two auxiliary matrices $P^{t}$ and $Z^{t}$
as 
\end{subequations}
\begin{equation}  \label{eq:TGA_APZ}
A^{t}=\frac{i}{2\hbar}\, P^{t}\cdot\left(Z^{t}\right)^{-1}.
\end{equation}
Since this decomposition is clearly not unique, a further constraint is
imposed, namely 
\begin{equation}
\dot{Z^{t}}=G\cdot P^{t}.  \label{eq:TGA_AZ_definition}
\end{equation}
In matrix notation, the unique solution of Eqs.~(\ref{eq:TGA_APZ}) and (\ref%
{eq:TGA_AZ_definition}) can be written as 
\begin{equation}  \label{eq:TGA_PZ_solution}
\renewcommand{\arraystretch}{0.5} \left(%
\begin{array}{c}
P^{t} \\ 
Z^{t}%
\end{array}%
\right)=M^{t}\cdot\left(%
\begin{array}{c}
P^{0} \\ 
Z^{0}%
\end{array}%
\right),
\end{equation}
with initial conditions $Z^{0}=I$ and $P^{0}=2i\hbar\, A^{0}$. The
time-dependent matrix $M^{t}:=\partial x^{t}/\partial x^{0}$ is the
stability matrix corresponding to the evolving phase-space point $x^{t}$.
Finally, by inserting Eqs.~(\ref{eq:TGA_APZ}) and (\ref{eq:TGA_PZ_solution})
into Eq.~(\ref{eq:TGA_equation_of_motions}), and by employing the matrix
identity $\det\exp B=\exp\mathrm{Tr}\,B$, one obtains directly an explicit
solution for $\gamma^{t}$ in the form 
\begin{equation}  \label{eq:TGA_gamma}
\gamma^{t}=\intop_{0}^{t}\mathcal{L}_{\mathrm{eff}}^{\tau}\, d\tau+\frac{%
i\hbar}{2}\ln(\det\, Z^{t}).
\end{equation}
Note that since the matrix $Z^{t}$ is complex, one has to ensure that a
proper branch of the logarithm be taken in order to make $\gamma^{t}$
continuous in time. 
%-------------------------------------------------------------------------------

\subsection{\label{subsec:OTF_TGA}On-the-fly \textit{ab initio\/} TGA}

%-------------------------------------------------------------------------------
In an OTF-AI (or ``direct'') dynamics, the required potential energy surface
is generated consecutively at each propagation step by any of the standard
electronic structure packages (see Sec.~\ref{sec:comp_details}). In addition
to classical trajectory propagation based only on force evaluation, TGA
requires to repeatedly evaluate the Hessian $\nabla^{2}V$ along the evolving
trajectory, since $\nabla^{2}V$ is needed\cite{Brewer_Manolopoulos:1997} for
propagating the stability matrix $M^{t}$.

The evolving GWP is properly defined only in the subspace of the vibrational
degrees of freedom of the molecule of interest. Therefore, a germane choice
of the coordinate system is essential. We illustrate the procedure employed
in the numerical calculations on a specific scenario of two, ground and
excited PESs, where the initial GWP corresponding to the ground vibrational
state of the excited electronic PES is subsequently propagated on the ground
electronic surface. Let us consider a reference equilibrium geometry $\xi _{%
\mathrm{ref}}$ on the excited PES, where $\xi _{\mathrm{ref}}$ is a vector
with $3N$ Cartesian components, with $N$ denoting the number of atoms in the
molecule. Any displaced molecular configuration $\xi $, obtained, e.g., by
propagation on a different PES, can be related to the normal-mode
coordinates $\eta $ as 
\begin{equation}
\xi -\xi _{\text{ref}}=G^{\frac{1}{2}}\cdot O\cdot \eta =T\cdot \eta ,
\label{eq:Cartesian_to_normalModes}
\end{equation}%
with $T:=G^{\frac{1}{2}}\cdot O$ and $O$ denoting the orthogonal matrix that
diagonalizes the mass-scaled Cartesian Hessian matrix evaluated at $\xi _{%
\mathrm{ref}}$, i.e., $T^{\mathsf{T}}\cdot \nabla ^{2}V\rvert _{\xi _{\text{%
ref}}}\cdot T=\Omega ^{2}$, where 
$\Omega =\rm{diag}(\omega_{1},\ldots,\omega_{3N})$ 
is the diagonal matrix containing the normal-mode
frequencies. Note that $\eta $ in Eq.~(\ref{eq:Cartesian_to_normalModes})
has $3N$ components, i.e., incorporates also the 3 translational and 3
rotational degrees of freedom. The initial values of these displacements are
zero and one would like to preserve this constraint also during the dynamics
on the ground PES. The translational modes are projected out easily by
shifting the center of mass to the origin of the Cartesian frame. Next, in
oder to minimize the coupling to the remaining 3 rotational modes, we
closely follow the axis-switching procedure devised by Hougen and Watson.%
\cite{Hougen_Watson:1965,Ozkan:1990} In this spirit, any displaced
configuration $\xi $ is rotated relatively to $\xi _{\mathrm{ref}}$ in order
to satisfy Eckart's conditions: 
\begin{equation}
\underset{a=1}{\sum^{N}}m_{a}\,(\mathrm{P}_{a}\cdot \xi _{\text{ref}})\times
\lbrack \mathrm{P}_{a}\cdot (\Lambda \cdot \xi )]=0.
\label{eq:Eckart_conditions}
\end{equation}%
In Eq.~(\ref{eq:Eckart_conditions}), the sum runs over all $N$ atoms, $%
\Lambda $ is a $3N\times 3N$ block-diagonal matrix, where each of the $N$
blocks is a copy of a $3$-dimensional rotation matrix $R$, and the $3\times
3N$ matrix $\mathrm{P}_{a}$ is defined as $(\mathrm{P}_{a})_{i,j}=\delta
_{i,1}\delta _{j,3a-2}+\delta _{i,2}\delta _{j,3a-1}+\delta _{i,3}\delta
_{j,3a}$. Application of $\mathrm{P}_{a}$ to a configuration $\xi $
essentially selects coordinates of the $a$th atom. Having minimized the
coupling to the rotational modes, one can afford to consider in Eq.~(\ref%
{eq:Cartesian_to_normalModes}) only the first $(3N-6)$ columns of the matrix 
$O$. In that case, the transformation matrix $T$ also reduces to a $3N\times
(3N-6)$ form. Kudin and Dymarsky showed\cite{Kudin_Dymarsky:2005} that the
rotation matrix $R$ solving Eq.~(\ref{eq:Eckart_conditions}) can be obtained
by minimizing the mass-weighted root-mean-square distance of $\xi $ with
respect to the reference configuration $\xi _{\mathrm{ref}}$. In practice,
this is achieved efficiently, e.g., by employing direct methods based on
singular value decomposition or quaternion formalism. \cite%
{Kearsley:1989,*Coutsias:2004}

The transformation from the Cartesian to the vibrational normal-mode
coordinates is thus performed in three consecutive steps. First, the
configuration $\xi$ is shifted to the center-of-mass system. Second, it is
rotated to the Eckart frame, and finally it is projected onto the
vibrational normal modes, i.e., 
\begin{equation}  \label{eq:cart2nm}
\eta = W\cdot\left[\Lambda\cdot\left(\xi-\Delta\right)-\xi_{\text{ref}}%
\right],
\end{equation}
where $W:=T^{\mathsf{T}}\cdot G^{-1}$ and the center-of-mass vector $\Delta$
is defined as 
\begin{equation}  \label{eq:cms}
\Delta:=\left(\sum_{a=1}^{N}\mathrm{P}_{a}\right)^{\mathsf{T}%
}\cdot\sum_{a=1}^{N}m_{a}\, \mathrm{P}_{a}\cdot\xi/\sum_{a=1}^{N}m_{a}.
\end{equation}
Finally, one also needs to express the Cartesian force $\nabla_{\xi}V$ and
the Cartesian Hessian matrix $\nabla_{\xi}^{2}V$ in the $\eta$\nobreakdash%
-coordinates: 
\begin{subequations}
\begin{align}
\nabla_{\eta}V &= (W\cdot\Lambda)\cdot\nabla_{\xi}V,  \label{eq:grad_trans}
\\
\nabla_{\eta}^{2}V &=
(W\cdot\Lambda)\cdot\nabla_{\xi}^{2}V\cdot(W\cdot\Lambda)^{\mathsf{T}}.
\label{eq:hess_trans}
\end{align}
%-------------------------------------------------------------------------------

\subsection{Stability matrix propagation: Symplecticity and effect of
Hessian interpolation}

%-------------------------------------------------------------------------------
The GWP's center and the accompanying stability matrix $M^{t}$ are
propagated classically using the second-order symplectic algorithm.\cite%
{Brewer_Manolopoulos:1997} Propagation of $M^{t}$ is the most expensive part
of the entire OTF-AI calculation since it requires knowledge of the Hessian
of the PES along the evolving trajectory.

The associated computational costs can be alleviated by employing a Hessian
update scheme, within which the Hessian is evaluated directly only once
every $s\geq1$ steps and approximated at the remaining steps with an
extrapolation method requiring gradients. Note that these Hessian update
schemes are in the context of dynamics typically used for the propagation of
the classical trajectory itself, e.g., within the framework of higher-order
predictor-corrector schemes (see Refs.~%
\onlinecite{Bakken:1999,*Hratchian:2005} and %
\onlinecite{Millam:1999,*Lourderaj:2007,*Wu_Zhuang:2010} and references
therein). In contrast, in Refs.~\onlinecite{Ianconescu:2013}, %
\onlinecite{Ceotto_Hase:2013}, and \onlinecite{Zhuang_Ceotto:2013} as well
as in the present work, approximative treatment of the Hessian is used only
for the propagation of $M^{t}$. However, whereas in Refs.~%
\onlinecite{Ianconescu:2013}, \onlinecite{Ceotto_Hase:2013}, and %
\onlinecite{Zhuang_Ceotto:2013} the Hessian update is based on
extrapolation, in this work polynomial interpolation of order $b$ is used to
obtain the Hessian at intermediate steps. A Hessian extrapolation update
scheme would be convenient in cases for which analytical \textit{ab initio}
Hessians are not available, e.g., for absorption spectrum calculation.

Note that our approach requires propagating the full classical trajectory
and storing the necessary information regarding the potential first, and
interpolating the Hessian later. The TGA GWP is computed in the second pass
through the stored data. The advantage of this approach is twofold: First,
the independent Hessian calculations in the second pass are easily
parallelized. Second, one can perform a global analysis of the trajectory
over the entire propagation range (see Subsec.~\ref%
{subsec:nn_analysis_theory}).

As the first test, we check the conservation of the \textit{symplectic
condition} 
\end{subequations}
\begin{equation}
{M^{t}}^{\mathsf{T}}\cdot J\cdot M^{t}=J  \label{eq:symplectic_conservation}
\end{equation}%
by the $2D\times 2D$ stability matrix $M^{t}$, where J is the standard
symplectic matrix 
\begin{equation*}
\renewcommand{\arraystretch}{0.9}J:=\left( 
\begin{array}{rr}
0_{D} & I_{D} \\ 
-I_{D} & 0_{D}%
\end{array}%
\right) 
\end{equation*}%
and $I_{D}$ is the $D$-dimensional identity matrix. The deviation from Eq.~(%
\ref{eq:symplectic_conservation}) is evaluated in terms of the error 
\begin{equation}
\epsilon ^{t}:=\Vert {M^{t}}^{\mathsf{T}}\cdot J\cdot M^{t}-J\Vert _{\text{F}%
},  \label{eq:symplecticity_of_M}
\end{equation}%
where $\Vert A\Vert _{\text{F}}:=\sqrt{\rm{Tr}(A^{\mathsf{T}}\cdot A)}=%
\sqrt{\sum_{i=1}^{2D}\sum_{j=1}^{2D}|A_{ij}|^{2}}$ denotes the Frobenius norm%
\cite{book:GolubLoan} of matrix $A$ and the exact stability matrix satisfies 
\begin{equation}
\epsilon ^{t}=0.  \label{eq:symplectic_cond}
\end{equation}%
For instance, in the T$2$ calculation, Eq.~(\ref{eq:symplectic_cond}) is
well satisfied even for Hessian interpolated from AI values computed only
every 2, 4, 8, or 16 steps. For details see Fig.~1(a) in the Supplementary
Material.\cite{supp_mat} It is important to note that any violations of Eq.~(%
\ref{eq:symplectic_cond}) are due to round-off errors, since in an
infinite-precision arithmetics, Eq.~(\ref{eq:symplectic_cond}) would be
satisfied even if the true Hessian were replaced by an arbitrary symmetric
matrix $\mathcal{H}^{t}$. The only additional requirement is that $M^{t}$ be
propagated symplectically,\cite{Brewer_Manolopoulos:1997} since the symmetry
of $\mathcal{H}^{t}$ is guaranteed by the interpolation algorithm.
Incidentally, note that Eq.~(\ref{eq:symplectic_cond}) is much more
stringent than the widely used Liouville condition, which only requires
conservation of the phase-space volume, expressed by the requirement $%
\mathop{\text{det}}({M^{t}}^{\mathsf{T}}\!\cdot M^{t})=1$, and automatically
follows from symplecticity [Eq.~(\ref{eq:symplectic_cond})].

The influence of the interpolation procedure with $s>1,\,b\geq0$ on the GWP
evolved with the TGA is quantified in terms of fidelity---a quantity
introduced by Peres\cite{Peres:1984} to measure sensitivity of quantum
dynamics to perturbations. In our setting, the fidelity is defined as the
squared magnitude of the time-dependent overlap of GWPs propagated using the
TGA with and without interpolation: 
\begin{equation}  \label{eq:F_sb}
F_{s,b}(t):=\left|\bigl\langle\psi_{1,b}^{t}|\psi_{s,b}^{t}\bigr\rangle%
\right|{}^{2}.
\end{equation}
In the T$2$ case, e.g., interpolating every four steps using the second
order interpolation ($b=2$) has almost no effect on the propagated GWP,
while the OTF-AI calculation is accelerated almost four times [see Fig.~1(b)
in the Supplementary Material\cite{supp_mat}]. 
%-------------------------------------------------------------------------------

\subsection{\label{subsec:nn_analysis_theory}Identification of the essential
DOFs}

%-------------------------------------------------------------------------------
Perhaps the greatest advantage of trajectory-based methods is the
possibility to visualize the dynamics and directly study its influence on
the resulting spectra. However, direct analysis can become quite cumbersome
for systems of high dimensionality. Moreover, the dynamical couplings among
individual degrees of freedom (DOFs) pose additional complications since all
the coupled DOFs must be analyzed simultaneously. In this subsection, we
introduce a particular approach for extracting the essential characteristics
of the dynamics of a generic system with $D$ vibrational DOFs. To some
extent, this method shares common grounds with other ``effective modes''
techniques aspiring to identify the modes responsible for the main spectral
features, e.g., methods tailored for the description of nonadiabatic
transitions.\cite{Cederbaum:2005,*Picconi:2013} However, in contrast to Ref.~%
\onlinecite{Cederbaum:2005,*Picconi:2013}, the identification of the
essential DOFs is here performed on the fly. The ``tool'' proposed here is
used in Subsec.~\ref{subsec:nn_analysis_results} for analyzing and
interpreting the emission spectra of the oligothiophene T$n$ family.

In order to simplify the discussion below, we introduce the symbol $\mathcal{%
D}$ to denote the space spanned by all $D$ DOFs. Any subspace of $\mathcal{D}
$ is then identified with the subset of indexes of those DOFs that span the
given subspace. In this spirit, $\mathcal{D}$ itself is identified with the
set $\mathcal{D}=\{1,\,2,\ldots,\,D\}$. Note that the set of normal mode
coordinates provide a natural physical realization of $\mathcal{D}$,
nevertheless our approach is not limited to this particular choice.

Briefly put, our strategy is as follows. First, we decompose the set $%
\mathcal{D}$ of all vibrational DOFs into mutually disjoint subsets, where
the DOFs in different subsets can be thought of as approximately dynamically
independent. Second, we identify the dynamically most important DOFs and
then consider only those subsets of $\mathcal{D}$ which contain at least one
of these ``important'' DOFs.

To quantify the coupling between various DOFs, we utilize the stability
matrix to measure the \textit{information flow} among individual DOFs. The
\textquotedblleft flow\textquotedblright\ $B_{ij}$ between $i$th and $j$th
DOF is then defined as 
\begin{equation}
B_{ij}:=\biggl|\frac{\beta _{ij}}{\beta _{ii}}\biggr|,\text{ with }\beta :=%
\frac{1}{T}\intop_{0}^{T}dt\,\kappa ^{\mathsf{T}}\cdot \tilde{M}^{t}\cdot
\kappa ,  \label{eq:analysis_B}
\end{equation}%
where $\tilde{M}_{ij}^{t}=\lvert M_{ij}^{t}\rvert $ and $\kappa ^{\mathsf{T}%
}=(I_{D},\,I_{D})$ denotes a two-component vector, each component of which
is a $D$-dimensional identity matrix. The value of $\beta _{ij}$ is rescaled
in Eq.~(\ref{eq:analysis_B}) by $1/\beta _{ii}$ in order to make the
diagonal elements unital ($B_{ii}=1$), as in uncoupled systems, and to focus
solely on the coupling effects among different DOFs ($i\neq j$). For
connection to and comparison with the analysis based on the global harmonic
model or Duschinsky matrix, we refer to Sec.~G of the Supplementary Material.%
\cite{supp_mat}

The decomposition of $\mathcal{D}$ into (approximately) dynamically
decoupled subsets of DOFs is then constructed by means of the concept of 
\textit{$\varepsilon_{B}$-partitioning}: 
\begin{equation}  \label{eq:eps_decomp}
\mathcal{D} = \bigcup_{\alpha=1}^{\bar{c}(\mathcal{D},\,\varepsilon_{B})}%
\mathcal{D}^{\varepsilon_{B}}_{\alpha},
\end{equation}
where $\bar{c}(\mathcal{D},\,\varepsilon_{B})$ denotes the number of
mutually disjoint subsets $\mathcal{D}^{\varepsilon_{B}}_{\alpha}$ defined
as the maximal connected components of an undirected graph with adjacency
matrix\cite{book_Chartrand} 
\begin{equation}
E_{ij} := \biggl\{%
\begin{array}{ll}
1 & \text{if }\max\bigl\{B_{ij},\,B_{ji}\bigr\}\geq\varepsilon_{B}, \\%
[1.5ex] 
0 & \text{otherwise},%
\end{array}%
\end{equation}
with a particular threshold value $\varepsilon_{B}$.

Any nontrivial decomposition (\ref{eq:eps_decomp}), where each subset $%
\mathcal{D}^{\varepsilon_{B}}_{\alpha}$ is interpreted as uncoupled, yields
a partially separable dynamics. Depending on the value $\bar{c}(\mathcal{D}%
,\,\varepsilon_{B})$, this separation can significantly reduce computational
costs, since the total correlation function can be obtained as a product of
individual contributions evaluated independently on each subspace (i.e.,
subset $\mathcal{D}^{\varepsilon_{B}}_{\alpha}$).

Next, we identify the set $\mathcal{G}^{\varepsilon_{\varrho}}\subseteq%
\mathcal{D}$ of the dynamically most important DOFs. For this purpose, we
employ the \textit{relative displacement} vector $\varrho$, the $i$th
component of which is defined as the maximal relative displacement in the
coordinate $\varsigma_{i}$, describing the $i$th DOF, where the maximum is
understood to be taken over the total propagation range $[0,\,T]$, i.e., 
\begin{equation}  \label{eq:analysis_rho}
\varrho_{i}:=\max_{0\leq t\leq T}\bigl|\varsigma_{i}^{t}\bigr| %
(A^{0}_{ii}/\ln\,2)^{1/2} \text{ for }1\leq i\leq D.
\end{equation}
Here, the scaling factor containing the diagonal element $A^{0}_{ii}$ of the
width matrix of the initial GWP ensures that the spread of the nuclear wave
function be taken into account: A small displacement of a high-frequency
(stiff) mode modulates the correlation function much more than the same
displacement of a low-frequency (soft) mode. The set $\mathcal{G}%
^{\varepsilon_{\varrho}}$ of dynamically most important modes is then
defined by 
\begin{equation}  \label{eq:G_rule}
i\in\mathcal{G}^{\varepsilon_{\varrho}}\Leftrightarrow\varrho_{i}\geq%
\varepsilon_{\varrho},
\end{equation}
where $\varepsilon_{\varrho}$ is a prescribed threshold value. A particular
DOF is thus interpreted as ``dynamically important'' if the dynamics
displaces it sufficiently relative to the width of the initial vibrational
state.

Finally, we combine the two ideas, i.e. the decoupling based on the $%
\varepsilon_{B}$-partitioning [Eq.~(\ref{eq:eps_decomp})], and the selection
of important modes based on the relative displacement $\varrho$ [Eq.~(\ref%
{eq:analysis_rho})], to form an ``active space'' $\mathcal{A}%
^{\varepsilon_{B},\,\varepsilon_{\varrho}}$ comprised of all subsets $%
\mathcal{D}_{\alpha}^{\varepsilon_{B}}$ containing at least one dynamically
important DOF from $\mathcal{G}^{\varepsilon_{\varrho}}$: 
\begin{equation}  \label{eq:active_space}
\mathcal{A}^{\varepsilon_{B},\,\varepsilon_{\varrho}} = \bigcup_{\alpha\in%
\mathcal{S}}\mathcal{D}^{\varepsilon_{B}}_{\alpha} \text{, with } \mathcal{S}%
:=\{\beta\!:\,\mathcal{G}^{\varepsilon_{\varrho}}\cap\mathcal{D}%
_{\beta}^{\varepsilon_{B}} \neq \emptyset\}.
\end{equation}
Note that the number $c(\mathcal{D},\,\varepsilon_{B},\,\varepsilon_{%
\varrho}):=\lvert\mathcal{S}\rvert$ of contributing subsets $\mathcal{D}%
^{\varepsilon_{B}}_{\alpha}$ is in general smaller than $\bar{c}(\mathcal{D}%
,\,\varepsilon_{B})$ of Eq.~(\ref{eq:eps_decomp}). In order to obtain a
contiguous labeling of the subsets in decomposition (\ref{eq:active_space}),
we introduce a bijective (but otherwise arbitrary) mapping $l$ between sets $%
\mathcal{S}$ and $\{1,\ldots,c(\mathcal{D},\,\varepsilon_{B},\,\varepsilon_{%
\varrho})\}$. This allows to restate Eq.~(\ref{eq:active_space}) as 
\begin{equation}  \label{eq:active_groups}
\mathcal{A}^{\varepsilon_{B},\,\varepsilon_{\varrho}} =\!\!\!\!
\bigcup_{\alpha=1}^{c(\mathcal{D},\,\varepsilon_{B},\,\varepsilon_{%
\varrho})}\!\!\mathcal{A}^{\varepsilon_{B},\,\varepsilon_{\varrho}}_{\alpha} 
\text{, where } \mathcal{A}^{\varepsilon_{B},\,\varepsilon_{\varrho}}_{%
\alpha}:=\mathcal{D}^{\varepsilon_{B}}_{l(\alpha)}.
\end{equation}
The subsets $\mathcal{A}^{\varepsilon_{B},\,\varepsilon_{\varrho}}_{\alpha}$
are in the following referred to as \textit{groups}. (Mathematically, these
``groups'' are just ``sets'' and should not be confused with a precise
mathematical notation of group.)

In summary, individual DOFs are by definition considered to be coupled only
within groups the union of which forms the space $\mathcal{A}%
^{\varepsilon_{B},\,\varepsilon_{\varrho}}$. Each group then contains at
least one DOF classified as dynamically important on the basis of the rule (%
\ref{eq:G_rule}). The total number of groups $c(\mathcal{D}%
,\,\varepsilon_{B},\,\varepsilon_{\varrho})$ and their structure is mainly
determined by the values of the two thresholds $\varepsilon_{\varrho}$ and $%
\varepsilon_{B}$ which have to be chosen appropriately according to the
system and process of interest.

\begin{figure*}[htp]
\centering
\includegraphics{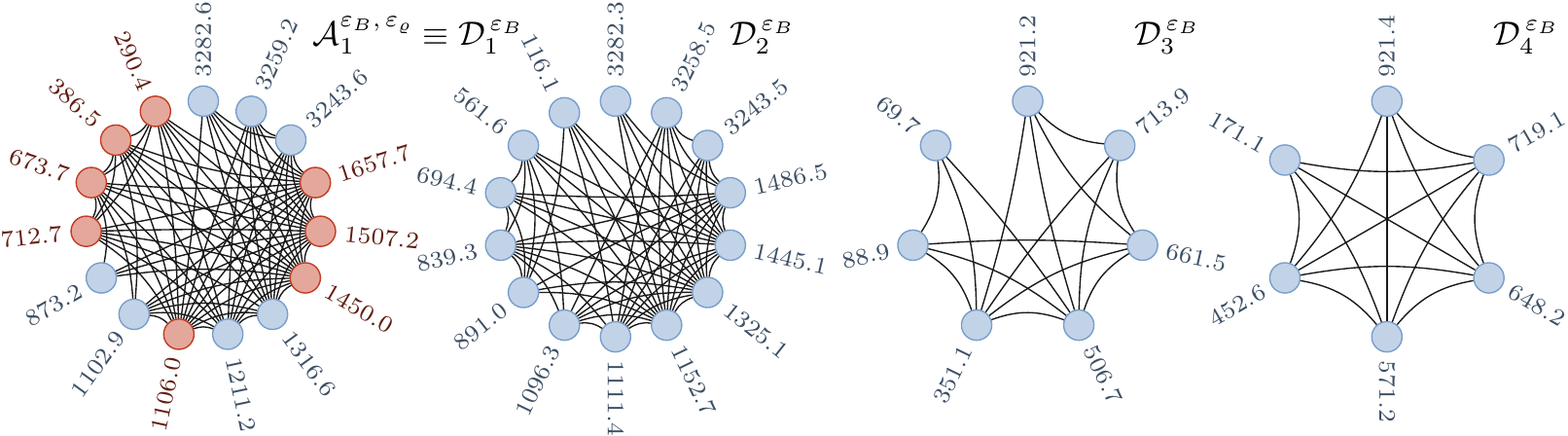}
\caption{Partitioning of $S_{1}$ normal-mode coordinates of dithiophene T$2$
into approximately independent subsets for the threshold value $\protect%
\varepsilon_{B}=0.045$ [see Eq.~(\protect\ref{eq:eps_decomp})]. Colored
circles represent individual modes, i.e., elements of $\mathcal{D}$. The
dynamically important modes [Eq.~(\protect\ref{eq:G_rule})] comprising $%
\mathcal{G}^{\protect\varepsilon_{\protect\varrho}}$ with the threshold
value $\protect\varepsilon_{\protect\varrho}=0.6$ are shown in red. Finally,
solid lines represent inter-mode couplings above the threshold $\protect%
\varepsilon_{B}$. Vibrational frequencies are given in cm$^{-1}$. }
\label{fig:epart_T2}
\end{figure*}

Let us now demonstrate the approach outlined above on one particular example
of T$2$, the dithiophene molecule. Since an oligothiophene T$n$ is comprised
of 
\begin{equation}  \label{eq:N_vs_n}
N(n)=7n+2
\end{equation}
atoms, the space $\mathcal{D}$ is of dimensionality $D(n):=3N(n)-6=21n$,
i.e., in the case of T$2$ ($n=2$), there are $42$ vibrational DOFs. To be
explicit, these vibrational DOFs are identified with normal-mode coordinates
of PES $S_{1}$. Individual modes are in Fig.~\ref{fig:epart_T2} represented
by colored circles with juxtaposed vibrational frequencies. Now, for $%
\varepsilon_{B}=0.045$, one obtains $4$ subsets in the decomposition (\ref%
{eq:eps_decomp}), i.e., $\bar{c}(\mathcal{D},\,\varepsilon_{B})=4$. Further,
we identify the set of important modes $\mathcal{G}$ using rule (\ref%
{eq:G_rule}). With threshold value $\varepsilon_{\varrho}=0.6$, we isolate $%
8 $ modes, i.e., $\lvert\mathcal{G}^{\varepsilon_{\varrho}}\rvert=8$. These
modes are shown in red color in Fig.~\ref{fig:epart_T2}. Finally, we see
that for this choice of the thresholds, we obtain only one group in the
decomposition (\ref{eq:active_groups}) since $\mathcal{G}^{\varepsilon_{%
\varrho}}\cap\mathcal{D}_{\beta}^{\varepsilon_{B}}\neq\emptyset$ only for $%
\beta=1$. Thus $c(\mathcal{D},\,\varepsilon_{B},\,\varepsilon_{\varrho})=1$
and the bijective mapping $l$ is merely an identity.

In practical calculations, $\varepsilon_{\varrho}$ and $\varepsilon_{B}$
must be chosen carefully. For high threshold values $\varepsilon_{B}$, one
can profit from an approximate separability of the model. However, too high
values of either $\varepsilon_{B}$ or $\varepsilon_{\varrho}$ might yield
inaccurate results. 
%-------------------------------------------------------------------------------
%-------------------------------------------------------------------------------

\section{\label{sec:comp_details}Computational details}

%-------------------------------------------------------------------------------
All \textit{ab initio} calculations were performed with the \texttt{%
Gaussian09} package.\cite{Gaussian09} Its output was extracted directly from
the checkpoint file. The ground PES $S_0$ was handled with the density
functional theory (DFT), whereas the first excited singlet PES ($S_1$) was
described with the time-dependent DFT (TD-DFT). Following the work of
Stendardo~\textit{et al.}, our TD-DFT calculations were based on the
long-range corrected CAM-B3LYP functional\cite{Stenardo:2012} with
6-31+G(d,p) basis set. Within this TD-DFT setup, the energy gap between the $%
S_0$ and $S_1$ PESs of oligothiophenes is described quite accurately.
Although \texttt{Gaussian09} provides analytical gradients for both DFT and
TD-DFT, analytical Hessians are available only for DFT. No symmetry
constraints were enforced and the ``fine'' and ``ultra fine'' integration
grids were used for OTF-AI calculations and geometry optimization,
respectively.

In order to find the physically relevant equilibrium geometry of $S_{1}$ for
each oligothiophene T$n$, we first performed an $S_0$ geometry optimization
of the ``all-trans'' conformer, the rings of which are oriented in an anti
conformation with respect to their neighbors. The work by Becker and
co-workers\cite{Becker:1996} suggests that this is the most stable
conformer. The $S_1$ geometry optimization was started from this $S_{0}$
equilibrium geometry. It has been well-established that in contrast to the
inter-ring twisted $S_{0}$ equilibrium geometry and its shallow potential, $%
S_{1}$ exhibits a steep, deep, harmonic-like well in the vicinity of its
planar equilibrium geometry.\cite{Belletete:1994,Becker:1996} The $S_{1}$
equilibrium geometry, shown in the Supplementary Material,\cite{supp_mat}
served as the reference structure for the OTF-AI-TGA dynamics.

Within the OTF-AI-TGA, the GWP was propagated for the total time of $7976\,%
\text{a.u.}\approx193\,\text{fs}$ with a time step of $8\,\text{a.u.}%
\approx0.2\,\text{fs}$ using the second order symplectic algorithm. The
resulting spectra were subjected to a phenomenological (inhomogeneous)
Gaussian broadening with half-width at half-maximum (HWHM) of $0.025$$\,%
\text{eV}\approx200\,\text{cm}^{-1}$. 
%-------------------------------------------------------------------------------

\section{\label{sec:results}Results and discussion}

%-------------------------------------------------------------------------------

\subsection{\label{subsec:experimental_spectrum}Comparison with experimental
spectra}

\begin{figure}[htp]
\centering
\includegraphics{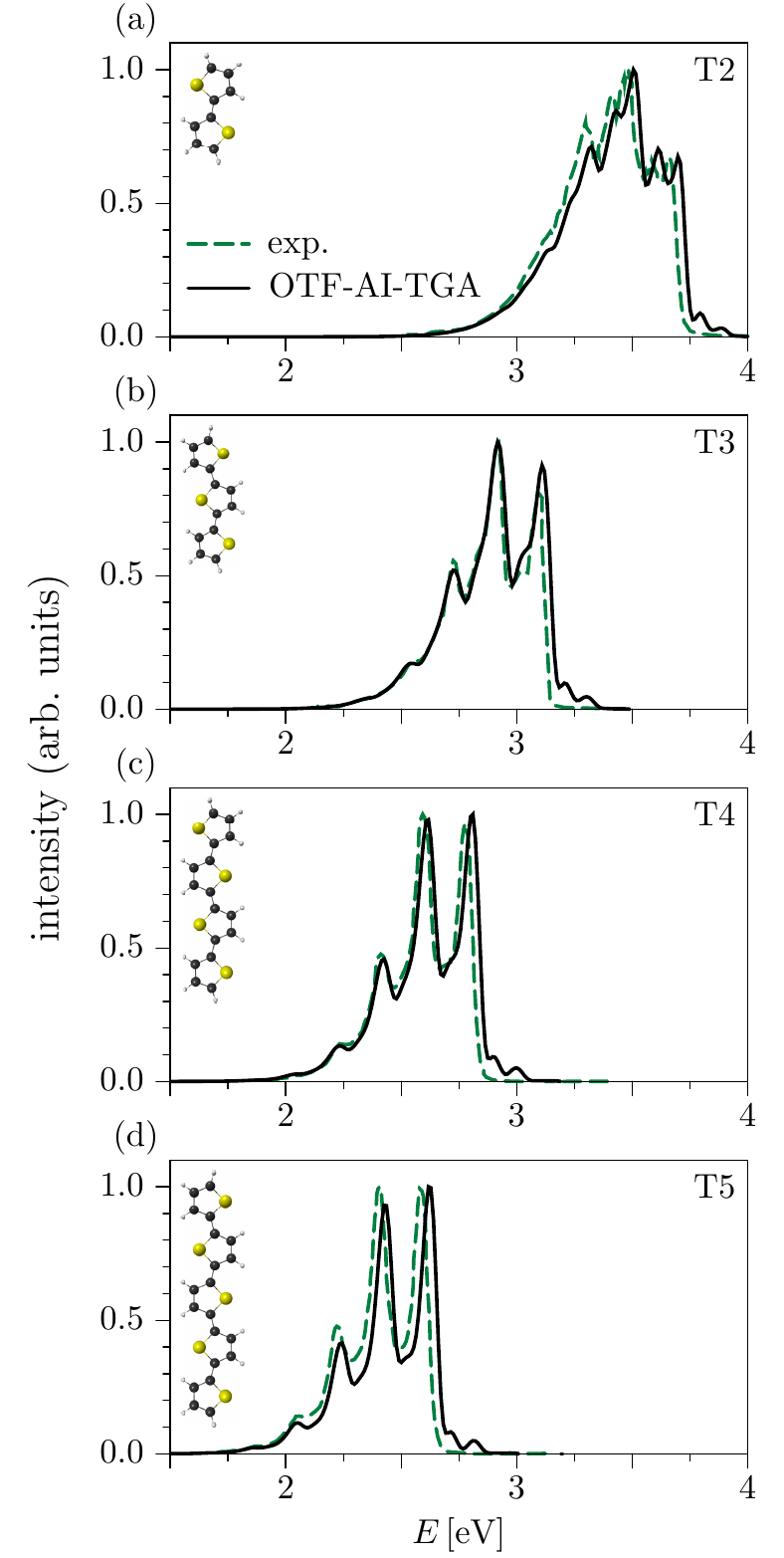} 
\caption{\label{fig:tga_vs_exp}
Emission spectra of the oligothiophene T$n$ family for $n\in\{2,3,4,5\}$: Comparison of experimental
emission spectra (exp., dashed green line) with the full-dimensional OTF-AI-TGA calculations using all 21$n$ normal modes (solid black line).
}
\end{figure}

Our results confirm the utility of the OTF-AI-TGA approach for electronic
spectra calculation, since all important features of the experimental
spectra are almost perfectly reproduced. Figure~\ref{fig:tga_vs_exp}
demonstrates the agreement with the overall shape, peak intensities, as well
as the trend of the spectra to gradually shift towards lower frequencies
with increasing number of rings in the molecule. Note that particular
experimental conditions, notably the interaction with the solvent (here,
ethanol glass at $77\,$K), can produce a shift of the spectrum. However, we
disregard this effect since the resulting shift is expected to be small for
a broad class of solvents.\cite{Becker:1996,Belletete:1994,Chadwick:1994}
Also, the exact prediction of the spectrum position is partly beyond the
level of the \textit{ab initio} setup employed here (see Sec.~\ref%
{sec:comp_details}).

%-------------------------------------------------------------------------------

%-------------------------------------------------------------------------------

Becker~\textit{et al.\/}\cite{Becker:1995}~reported a significant red shift
of the oligothiophene absorption spectra at low temperatures and attributed
this phenomenon to the twisted$\mapsto$planar conformational transition
induced by solvent freezing. Interestingly, this shift was not observed in
the emission spectra, which suggests that in the whole temperature range it
is only the planar conformation that plays a significant role in this
process. Even without imposing explicit planarity constraints, no deviations
from the planar conformation were observed during the ground-state gas-phase
OTF-AI dynamics due to planarity of the initial geometry. This fact makes
the comparison of our gas-phase results to the experimental data more
legitimate. Finally note that the \textit{ab initio} ground state
equilibrium geometry is twisted in contrast to the equilibrium geometry in
ethanol glass at $77\,$K. Therefore, the $n-1$ torsional degrees of freedom
connecting the planar and twisted geometries of T$n$ have imaginary
frequencies. Since our approach is unable to describe wave-packet splitting,
the TGA GWP only spreads along these degrees of freedom (see Supplementary
Material,\cite{supp_mat} Sec.~E). However, since we are mainly interested in
short-time dynamics, this behavior is qualitatively correct. Hence, the
OTF-AI-TGA approach remains in this case robust even for floppy molecules
and the question about the ``harmonicity'' of the system is of much lesser
importance due to the employment of the local harmonic approximation.
Although the global harmonic approximation is quite adequate for T$n$,\cite%
{Stenardo:2012} small changes of the peak positions and intensities can be
observed as compared to OTF-AI-TGA (see Supplementary Material,\cite%
{supp_mat} Sec.~F).

\begin{figure}[htp]
\centering
\includegraphics{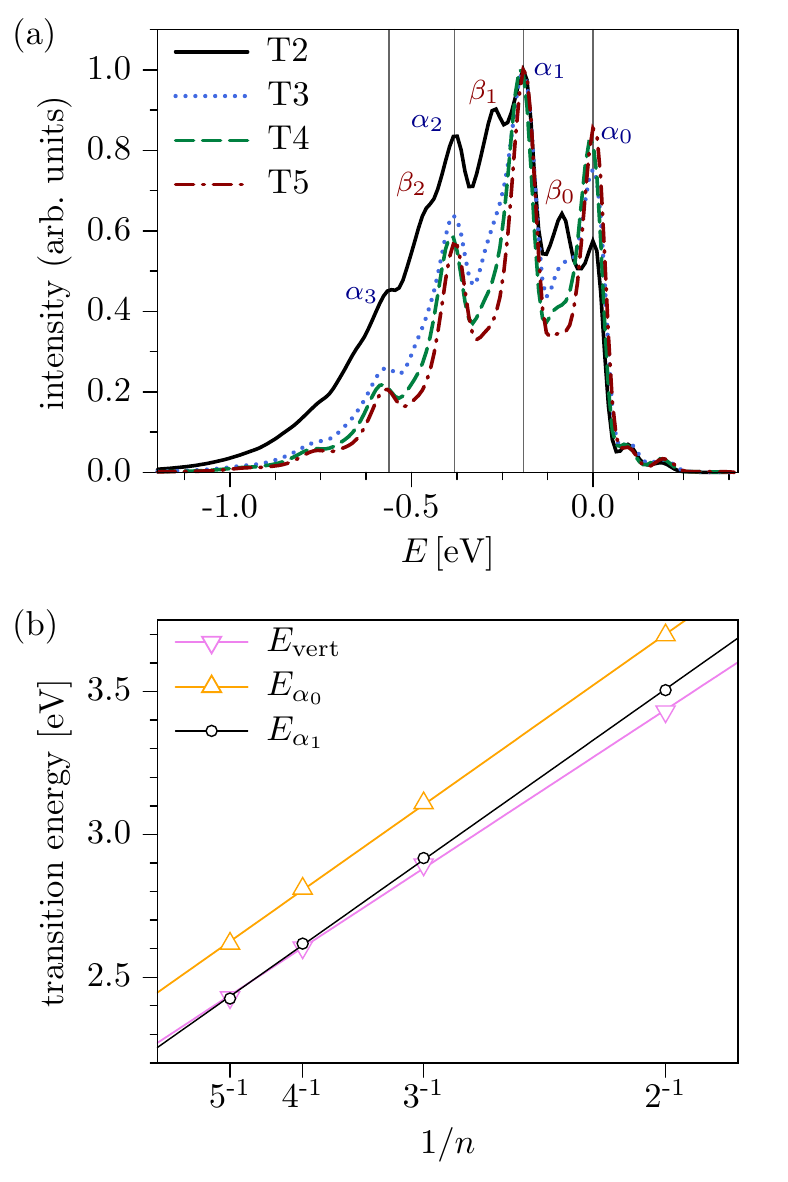} 
\caption{\label{fig:spec}
Emission in the oligothiophene T$n$ family for $n\in\{2,3,4,5\}$.
(a)~$L^{\infty}$-normalized line-shape spectra. To facilitate their comparison, the spectra are shifted
independently for each $n$ so that the $\alpha_{0}$ peak appears at zero energy.
(b)~Dependence of the vertical-transition energy $E_{\text{vert}}$ and positions of the $\alpha_{0}$ and $\alpha_{1}\text{-peaks}$
$(E_{\alpha_0},\,E_{\alpha_1})$ on $1/n$ (see text for details). Linear fits are denoted with lines.
}
\end{figure}

To facilitate comparison between line-shape spectra of oligothiophenes with
different numbers of thiophene rings, the spectra shown in Fig.~\ref%
{fig:spec}(a) are first $L^{\infty}$ normalized and subsequently shifted so
that the ``$\alpha_{0}$-peaks'' overlap at zero energy. This reveals that
the relative peak positions are rather insensitive to $n$, while their
prominence is increasing with increasing $n$. The peak at the highest energy
(in our notation: $\alpha_{0}$) in the emission spectrum is attributed to
the $0$--$0$ transition.\cite{Gierschner:2007} The position of the $%
\alpha_{1}$-peak is close to the vertical transition energy $E_{\text{vert}}$%
, which, in loose terms, justifies its dominance in Fig.~\ref{fig:spec}(a).
More detailed classification of individual spectral peaks into the $%
\alpha,\beta$ groups and their interpretation from the dynamical viewpoint
is discussed in Subsec.~\ref{subsec:nn_analysis_results}.

%-------------------------------------------------------------------------------

%-------------------------------------------------------------------------------
It has been found experimentally that the $0$--$0$ transition energy $E_{0%
\text{--}0}$ in the polythiophene family T$n$ is a linear function of $1/n$.%
\cite{Becker:1996,Yang:1997,Yang:1998,Themans:1989} In accordance with this
observation and our identification of $E_{0\text{--}0}$ with $E_{\alpha_0}$,
we found that $E_{\alpha_{0}}$ is accurately described by the function $%
E_{\alpha_{0}}(n)\approx(3.58/n+1.91)\,\text{eV}$. Good agreement with the
experiment can be directly inferred from Fig.~\ref{fig:tga_vs_exp}.
Furthermore, from the \textit{ab initio} data, we determined in a similar
fashion that $E_{\text{vert}}(n)\approx(3.33/n+1.77)\,\text{eV}$. Fits of $%
E_{\alpha_0}$, $E_{\alpha_1}$ and $E_{\text{vert}}$ are shown in Fig.~\ref%
{fig:spec}(b).

Note that the relative intensity of the $\alpha_{0}$-peak, identified with
the $0$--$0$ transition, in Fig.~\ref{fig:spec}(a) increases with $n$. This
can be related to the fact that the slope of $E_{\alpha_{1}}(n)$ is larger
than the slope of $E_{\text{vert}}(n)$ [see Fig.~\ref{fig:spec}(b)], using
the following heuristic argument: Neglecting the difference between the $%
S_{0}$ and $S_{1}$ zero-point energies, the $0$--$0$ transition energy
depends solely on the energy gap between these PESs. On the other hand, $E_{%
\text{vert}}$ is influenced also by the relative displacement of the $S_{0}$
and $S_{1}$ potential minima. Therefore, if $E_{\text{vert}}$ decreases more
slowly with increasing $n$ than does the $0$--$0$ transition energy, one can
expect a decrease not only in the energy gap between $S_{0}$ and $S_{1}$
PESs but also in the relative displacement of their minima, which, in turn,
is responsible for the gain in intensity of the $0$--$0$ transition, i.e.,
the $\alpha_{0}$-peak. This observation is in agreement with the Huang-Rhys
analysis performed by A.~Yang \textit{et al.}\cite{Yang:1998} on
fluorescence spectra of T$n$ for $n\in\{3,4,5,6\}$. 
%-------------------------------------------------------------------------------

\subsection{\label{subsec:nn_analysis_results}Vibrational analysis}

%-------------------------------------------------------------------------------
To gain a deeper understanding of the emission spectra shown in Figs.~\ref%
{fig:tga_vs_exp} and \ref{fig:spec}, we employ independently for each
oligothiophene T$n$ the analysis proposed in Subsec.~\ref%
{subsec:nn_analysis_theory} adapted to the normal-mode coordinates of the $%
S_{1}$ PES of T$n$. To this end, we closely follow the example presented at
the end of Subsec.~\ref{subsec:nn_analysis_theory}. The normal-mode
classification based on decompositions (\ref{eq:active_groups}) and (\ref%
{eq:classes}) with $\varepsilon_{B}=0.55$ and $\varepsilon_{\rho}=0.6$
results for all T$n$ in an active space $\mathcal{A}$ comprised of six
groups of modes (see Tab.~\ref{tab:mode_analysis}). This space is spanned by
ten ``active'' modes (i.e., $\lvert\mathcal{A}\rvert=10$) for T$2$-T$4$,
while $\lvert\mathcal{A}\rvert=8$ for T$5$. The thresholds were chosen in
order to obtain a minimal set $\mathcal{A}$ of active modes with as many
subsets as possible on condition that the reduced OTF-AI-TGA spectrum $%
\sigma_{\mathcal{A}}$ recovers all important features of the ``complete''
spectrum $\sigma_{\mathcal{D}}$. For clarity, the subscript of $\sigma$
denotes explicitly the set of modes taken into account in the spectra
calculation. Formally, the spectrum $\sigma_{\mathcal{A}}$ can be thought of
as the computationally cheapest, yet still sufficiently accurate
approximation of $\sigma_{\mathcal{D}}$.

For details regarding correlation function and spectra calculations within
proper subspaces of $\mathcal{D}$ we refer to the Appendix~\ref{sec:rd}.
From now on, to simplify notation, the implicit dependence of, e.g., $%
\mathcal{A}$ on the threshold values $\varepsilon_{B}$ and $%
\varepsilon_{\varrho}$ will not be denoted explicitly.

\begin{figure}[h!t]
\centering
\includegraphics{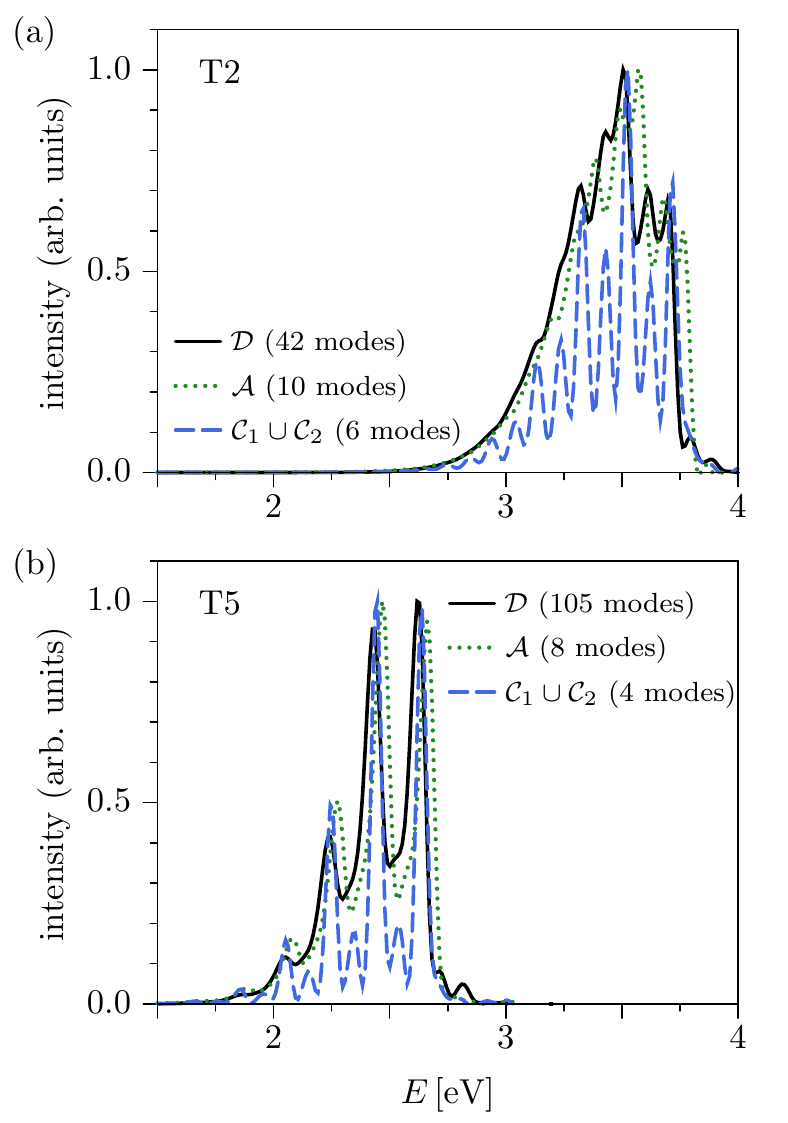} 
\caption{\label{fig:rd}
Emission spectra of oligothiophenes T$2\,$(a) and T$5\,$(b): comparison of the full-dimensional
OTF-AI-TGA spectrum $\sigma_{\mathcal{D}}$ (solid black line) with the spectrum $\sigma_{\mathcal{A}}$
(dotted green line) computed within the subspace $\mathcal{A}$ of the active modes
and the spectrum $\sigma_{\mathcal{C}_{1}\cup\,\mathcal{C}_{2}}$ (dashed blue line) taking into
account only modes belonging to the classes $\mathcal{C}_{1}$ and $\mathcal{C}_{2}$ (see Fig.~\ref{fig:spec_classes} and Tab.~\ref{tab:mode_analysis})
introduced in Eq.~(\ref{eq:classes}).
}
\end{figure}

Figure~\ref{fig:rd}(a) demonstrates that ten modes were sufficient to
essentially reproduce the complete spectrum $\sigma_{\mathcal{D}}$ for T$2$.
The simplification achieved is the most striking for T$5$ [Fig.~\ref{fig:rd}%
(b)], for which eight modes were sufficient and hence the dimensionality was
reduced more than ten times without losing any major feature in the
spectrum. However, note that the ``$\lvert\mathcal{A}\rvert$-mode'' spectra
in Fig.~\ref{fig:rd} are slightly shifted due to dependence of the
zero-point energy on the choice of $\mathcal{A}$. (Analogous spectra of T$3$
and T$4$ are shown in Sec.~B of the Supplementary Material.\cite{supp_mat}) 
%-------------------------------------------------------------------------------

%-------------------------------------------------------------------------------

\begin{figure*}[htp]
\centering
\begin{tabular}{p{2.0in}@{}p{1.9in}@{}p{2.4in}}
(a)~inter-ring stretch & (b)~ring squeeze & (c)~chain deformation\tabularnewline
\includegraphics{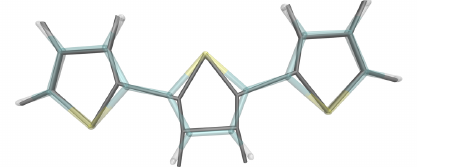} & \includegraphics{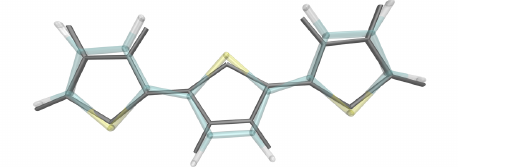} & \includegraphics{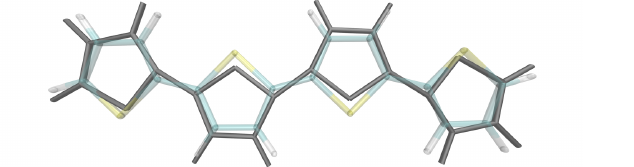}\tabularnewline[-1.5ex]
(d)~inner-ring C-S-C stretch & (e)~ring expansion & (f)~C-H deformation\tabularnewline
\includegraphics{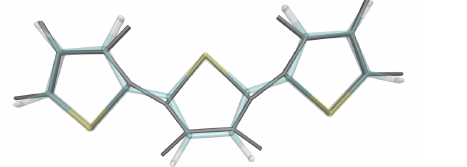} & \includegraphics{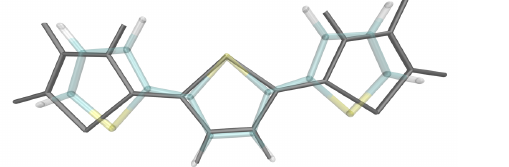} & \includegraphics{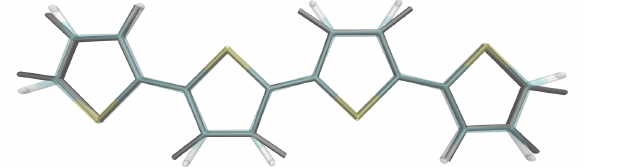}\tabularnewline[-1.5ex]
\multicolumn{3}{l}{(g)~C-S-C outer-ring asymmetric stretch} \tabularnewline
\includegraphics{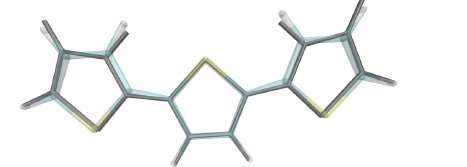} & &
\end{tabular}
\caption{Characterization of the active normal modes in the set $\mathcal{A}%
\subseteq\mathcal{D}$ [see Eqs.~(\protect\ref{eq:active_groups}) and (%
\protect\ref{eq:classes})] by the nature of the deformation which they exert
on the oligothiophene T$n$ skeleton. To cover all cases presented in Tab.~%
\protect\ref{tab:mode_analysis}, these deformations are shown on the
examples of T$3$ and T$4$ . Panel labels correspond to the classification in
Tab.~\protect\ref{tab:mode_analysis}.}
\label{fig:deformation_classes}
\end{figure*}

The modes in $\mathcal{A}$ are by definition considered to be coupled only
within individual groups. Therefore, one can attempt to assign a
characteristic vibrational movement of the entire molecule induced by
excitation of the modes belonging to a particular group. Among the $24$
groups ($24=4$ oligothiophenes $\times$ $6$ groups per oligothiophene), we
identified $7$ characteristic motions shown on the examples of T$3$ and T$4$
molecules in Fig.~\ref{fig:deformation_classes}. In Table~\ref%
{tab:mode_analysis}, these characteristic motions are distinguished with a
superscript. 
%-------------------------------------------------------------------------------
%-------------------------------------------------------------------------------
\begin{figure*}[htp]
\centering
\includegraphics{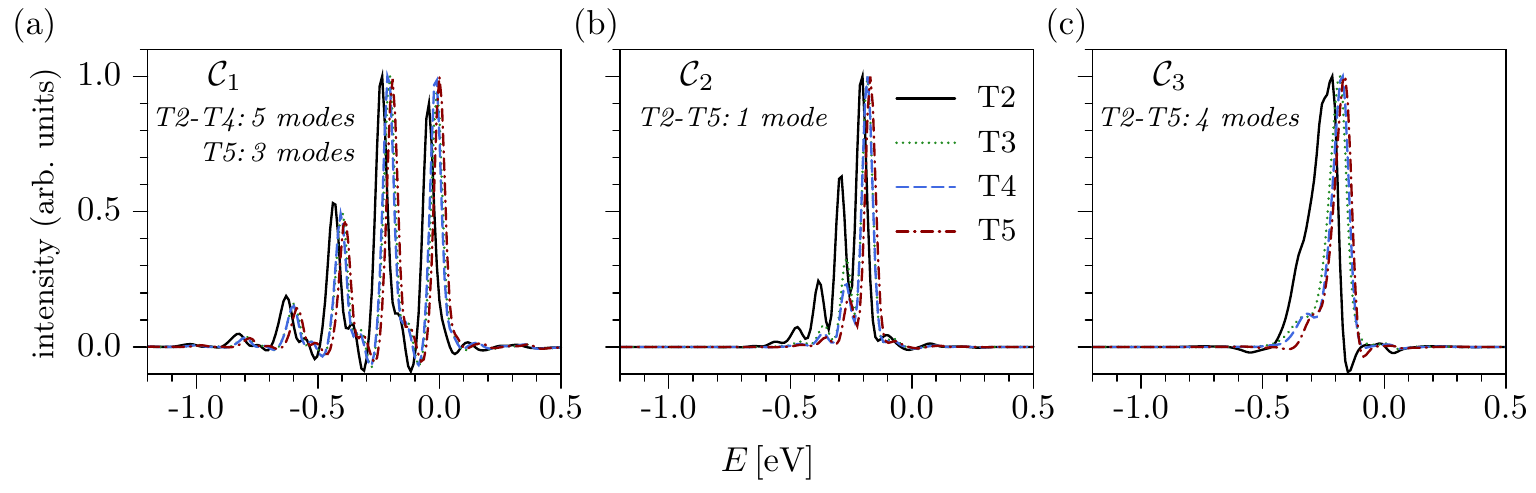}
\caption{Classification of normal modes of the oligothiophene T$n$ family
according to their influence on the resulting emission spectrum [see Eqs.~(%
\protect\ref{eq:active_groups}) and (\protect\ref{eq:classes})]. Detailed
description of individual classes is contained in Tab.~\protect\ref%
{tab:mode_analysis}. (a)~Inter-ring stretch modes responsible for the $%
\protect\alpha$-peaks shown in Fig.~\protect\ref{fig:spec}. (b)~Ring-squeeze
mode reflected in the $\protect\beta$-peaks in Fig.~\protect\ref{fig:spec}.
(c)~Remaining modes causing overall broadening of the spectra. }
\label{fig:spec_classes}
\end{figure*}
%-------------------------------------------------------------------------------

Next, the six groups of normal modes are, for each $n\in\{2,3,4,5\}$, merged
into three disjoint \textit{classes} $\mathcal{C}_{1}$, $\mathcal{C}_{2}$,
and $\mathcal{C}_{3}$ as 
\begin{equation}  \label{eq:classes}
\mathcal{C}_{1}:=\mathcal{A}_{1},\text{ } \mathcal{C}_{2}:=\mathcal{A}_{2},%
\text{ and } \mathcal{C}_{3}:=\bigcup_{\alpha=3}^{6}\mathcal{A}_{\alpha}.
\end{equation}
The reason for introducing an additional logical layer is the observation in
Fig.~\ref{fig:spec_classes} that the overall character of the spectrum $%
\sigma_{\mathcal{C}_{i}}$ corresponding to the $i$th group is only mildly
influenced by $n$, whereas the dependence on $i$ is dominant. In loose
terms, the first group $\mathcal{C}_{1}$ comprises inter-ring stretch modes
and is mainly reflected in the ``$\alpha$-peaks'' of the complete spectrum $%
\sigma_{\mathcal{D}}$ [see Fig.~\ref{fig:spec}(a)]. The second group $%
\mathcal{C}_{2}$ consists of a ring-squeeze mode and produces the ``$\beta$%
-peaks'' in Fig.~\ref{fig:spec}(a). Finally, the modes contained in the
third group cause merely an overall broadening of the spectrum. Such a
classification of vibrational modes, essential for a theoretical
interpretation of the emission spectra, is also useful in practice, e.g., in
the design of organic light-emitting diodes (OLEDs).\cite{Lumpi:2013}

The difference between individual classes is further emphasized by
introducing an ``overall relative displacement'' of the $i$th class as $%
R_{i}^{2}:=\sum_{j\in\mathcal{C}_{i}}\varrho_{j}^{2}$. We have found that $%
R_{1}$ is highly correlated with $-n$ while $R_{2}$ with $1/n$. Therefore,
for low $n$, the dynamical importance of the class $\mathcal{C}_{2}$
decreases faster with increasing $n$. This results in less structured
spectra, shown in Fig.~\ref{fig:spec}(b), in which the $\beta$-peaks are
almost invisible already for T$3$.

In summary, the inter-ring stretch motion is seen to have a dominant effect
on the T$n$ spectra, especially for $n>2$. Comparing the relative
displacements of the classes $\mathcal{C}_1$ and $\mathcal{C}_2$ helps to
further corroborate the hypothesis (stated above) that the $S_{0}$ and $%
S_{1} $ geometries become less displaced with increasing $n$ since the $0$--$%
0$ transition energy $E_{0\text{--}0}(n)$ decreases faster than the vertical
excitation energy $E_{\text{vert}}(n)$. 
%-------------------------------------------------------------------------------

\subsection{\label{subsec:quinoid_structure}Quinoid structure of $S_{1}$}

\begin{figure}[th!]
\centering
\includegraphics{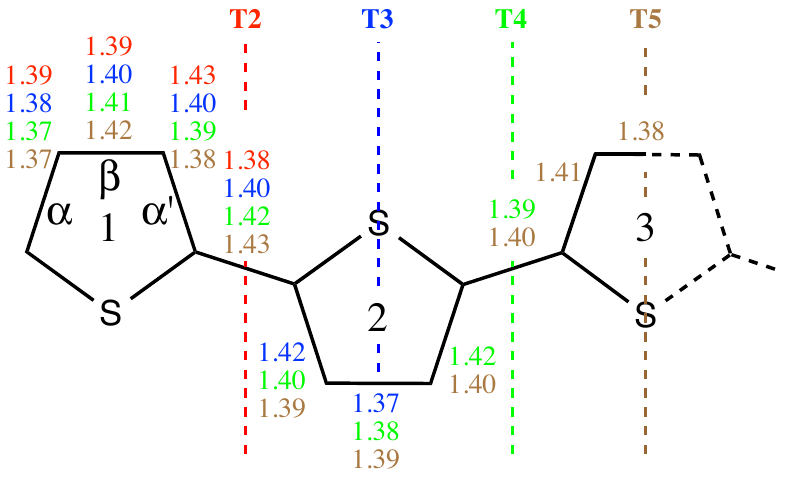} 
\caption{\label{fig:S1_polythiophene}
Equilibrium $S_{1}$ geometry of oligothiophenes T$n$ for $n\in\{2,3,4,5\}$. Corresponding bond lengths for different oligothiophenes T$n$
are juxtaposed with each other next to individual bonds, whereas
the dashed lines represent the end of the half-chain for each T$n$. E.g., to the right of the dashed line marked as T$3$ there are only one or two bond-length values  since those bonds are not present in the half-chain of T$2$ and T$3$.
}
\end{figure}

%-------------------------------------------------------------------------------
The extent of $\pi$-conjugation along the oligomer chain is reflected in the
quinoid structure of individual rings. The degree of the quinoid/aromatic
character of the $i$th ring in T$n$ can be quantified in terms of the
so-called \textit{bond length alternation}\cite%
{Beljonne:1996,Fazzi:2010,Ortiz:2010} (BLA) 
\begin{equation}  \label{eq:bla}
\text{BLA}_{i}=R_{\beta_{i}}-(R_{\alpha_{i}}+R_{\alpha_{i}^{\prime}})/2,
\end{equation}
where $R$ denotes the length of the $\beta$, $\alpha$, and $\alpha^{\prime}$
bonds of the $i$th ring (see Fig.~\ref{fig:S1_polythiophene}). Hence,
quinoid rings have a negative BLA, while aromatic rings have a positive BLA. 
%-------------------------------------------------------------------------------

%-------------------------------------------------------------------------------

The $S_{1}$ equilibrium geometries of T$n$ in Fig. \ref{fig:S1_polythiophene}
reveal that for $n>2$, both quinoid and aromatic ring types are present in
the chain: The inner rings are quinoid, while the end rings are aromatic. On
the other hand, both rings of T$2$ have quinoid character. However, the
large difference between the lengths of $\alpha$ and $\alpha^{\prime}$ bonds
suggests a double-bond character of the outer $\alpha$ bond in T${}_2$. In
general, the DFT $S_{1}$ geometries exhibit more pronounced quinoid
character in comparison with the $S_{1}$ geometries calculated at the MNDO
level,\cite{Beljonne:1996} which describe T$2$ as slightly aromatic.

The time dependence of BLA, displayed for T$5$ in Fig.~\ref{fig:bla} and for
T2, T3, and T4 in Sec.~C of the Supplementary Material,\cite{supp_mat} shows
emission-induced oscillations between the quinoid and aromatic characters of
individual rings. The inner rings are seen to be subjected to larger
structural variations, while the outer rings remain aromatic, although the
degree of aromaticity changes periodically. Hence, the quinoid character of T%
$n$ in $S_{1}$ is well localized over just $2$-$3$ rings, as was shown also
by Beljonne \textit{et al.},\cite{Beljonne:1996} while the emission process
triggers deformation of the whole chain.

Oligomer vibrational line shapes are usually analyzed in terms of the
effective conjugation coordinate\cite%
{Zerbi:1989,Geisselbrecht:1993,Zerbi:2007} (ECC)---a totally symmetric
internal coordinate describing the variation of adjacent C-C backbone
stretches, responsible for the change from the aromatic to quinoid
structure. A detailed analysis (summarized in Appendix~\ref{sec:ecc}) of the
dynamics shows that only some of the modes coupled to ECC are also excited
by the fluorescence process. The overall contribution of the $\mathcal{A}%
_{1} $ group to the ECC is more than $92$\% for all olighothiophenes and,
hence, the $\alpha$-peaks originate from the change of the ECC during the
dynamics induced by the fluorescence process. 
%-------------------------------------------------------------------------------

\begin{figure}[th!]
\centering
\includegraphics{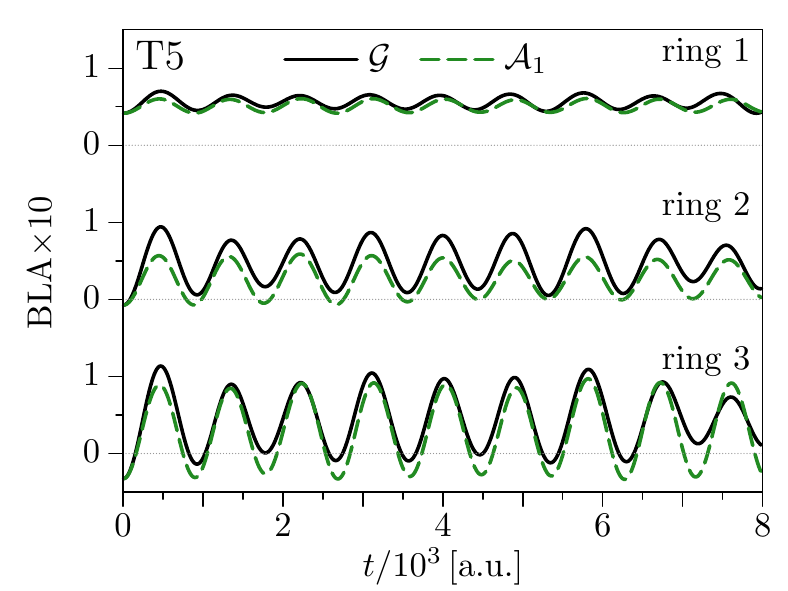} 
\caption{\label{fig:bla}
Time dependence of the bond length alternation (BLA) parameter 
during the dynamics induced by the emission in pentathiophene T$5$ [see Eq.~(\ref{eq:bla}) and Fig.~\ref{fig:S1_polythiophene}]. The character of the
outer rings (rings~$1$ and $2$) is mainly aromatic (positive BLA), while the transition to
the quinoid structure (negative BLA) occurs almost exclusively within the inner ring (ring $3$).
}
\end{figure}

%-------------------------------------------------------------------------------

%-------------------------------------------------------------------------------
\begin{table*}[htp]
\caption{Normal-mode classification based on decompositions (\protect\ref%
{eq:active_groups}) and (\protect\ref{eq:classes}) with $\protect\varepsilon%
_{B}=0.55$ and $\protect\varepsilon_{\protect\rho}=0.6$ for the
oligothiophene T$n$ family, $n\in\{2,3,4,5\}$. Vibrational frequencies $%
\protect\omega_i$ are given in cm$^{-1}$, while the maximum relative
displacements $\protect\rho_{i}$ of Eq.~(\protect\ref{eq:analysis_rho}) are
dimensionless. The modes are further classified into $7$ groups by the
character of the deformation which they exert on the oligothiophene
skeleton. These groups are distinguished by superscript labels next to
frequency values. For schematic depiction of these deformations see Fig.~%
\protect\ref{fig:deformation_classes}.}
\label{tab:mode_analysis}\renewcommand{\arraystretch}{1.0} %
\xdefinecolor{seda}{RGB}{207,207,196} \xdefinecolor{zelena}{RGB}{119,221,119}
\xdefinecolor{modra}{RGB}{174,198,207} %
\xdefinecolor{cervena}{RGB}{255,105,97} %
\xdefinecolor{oranzova}{RGB}{255,179,71} %
\xdefinecolor{zluta}{RGB}{253,253,150} 
\begin{tabular}{cc*{4}{D{.}{.}{5.4}@{}}*{4}{D{.}{.}{3.5}@{}}}
\midrule\midrule
\multirow{2}{*}{class}&\multirow{2}{*}{group}&%
\multicolumn{4}{c}{$\omega_i\,$[cm$^{-1}$]}&%
\multicolumn{4}{c}{$\rho_i$}\\[-1ex]
&&%
\multicolumn{1}{c}{T$2$}&%
\multicolumn{1}{c}{T$3$}&%
\multicolumn{1}{c}{T$4$}&%
\multicolumn{1}{c}{T$5$}&%
\multicolumn{1}{c}{T$2$}&%
\multicolumn{1}{c}{T$3$}&%
\multicolumn{1}{c}{T$4$}&%
\multicolumn{1}{c}{T$5$}\\\midrule
\multirow{5}{*}{$\mathcal{C}_{1}$}&%
\multirow{5}{*}{$\mathcal{A}_{1}$}&%
1657.7\mbox{${}^{\scriptscriptstyle(a)}$}&1630.5\mbox{${}^{\scriptscriptstyle(a)}$}&1615.5\mbox{${}^{\scriptscriptstyle(a)}$}&1598.6\mbox{${}^{\scriptscriptstyle(a)}$}&2.57\mbox{${}^{\scriptscriptstyle(a)}$}&2.50\mbox{${}^{\scriptscriptstyle(a)}$}&2.41\mbox{${}^{\scriptscriptstyle(a)}$}&1.93\mbox{${}^{\scriptscriptstyle(a)}$}\\
&&1507.2\mbox{${}^{\scriptscriptstyle(a)}$}&1553.9\mbox{${}^{\scriptscriptstyle(a)}$}&1545.2\mbox{${}^{\scriptscriptstyle(a)}$}&1590.1\mbox{${}^{\scriptscriptstyle(a)}$}&1.23\mbox{${}^{\scriptscriptstyle(a)}$}&1.10\mbox{${}^{\scriptscriptstyle(a)}$}&1.12\mbox{${}^{\scriptscriptstyle(a)}$}&1.35\mbox{${}^{\scriptscriptstyle(a)}$}\\
&&1450.0\mbox{${}^{\scriptscriptstyle(a)}$}&1501.2\mbox{${}^{\scriptscriptstyle(a)}$}&1581.3\mbox{${}^{\scriptscriptstyle(a)}$}&1548.5\mbox{${}^{\scriptscriptstyle(a)}$}&0.88\mbox{${}^{\scriptscriptstyle(a)}$}&0.86\mbox{${}^{\scriptscriptstyle(a)}$}&0.56\mbox{${}^{\scriptscriptstyle(a)}$}&1.13\mbox{${}^{\scriptscriptstyle(a)}$}\\
&&1211.2\mbox{${}^{\scriptscriptstyle(a)}$}&1461.9\mbox{${}^{\scriptscriptstyle(a)}$}&1498.2\mbox{${}^{\scriptscriptstyle(a)}$}& &0.59\mbox{${}^{\scriptscriptstyle(a)}$}&0.36\mbox{${}^{\scriptscriptstyle(a)}$}&0.55\mbox{${}^{\scriptscriptstyle(a)}$}&\\
&&3243.6\mbox{${}^{\scriptscriptstyle(a)}$}&1341.6\mbox{${}^{\scriptscriptstyle(a)}$}&1462.6\mbox{${}^{\scriptscriptstyle(a)}$}& &0.25\mbox{${}^{\scriptscriptstyle(a)}$}&0.34\mbox{${}^{\scriptscriptstyle(a)}$}&0.36\mbox{${}^{\scriptscriptstyle(a)}$}&\\\midrule
$\mathcal{C}_{2}$&%
$\mathcal{A}_{2}$&%
673.7\mbox{${}^{\scriptscriptstyle(b)}$}&696.7\mbox{${}^{\scriptscriptstyle(b)}$}&704.7\mbox{${}^{\scriptscriptstyle(b)}$}&710.8\mbox{${}^{\scriptscriptstyle(b)}$}&2.05\mbox{${}^{\scriptscriptstyle(b)}$}&1.56\mbox{${}^{\scriptscriptstyle(b)}$}&1.32\mbox{${}^{\scriptscriptstyle(b)}$}&1.17\mbox{${}^{\scriptscriptstyle(b)}$}\\\midrule
\multirow{4}{*}{$\mathcal{C}_{3}$}&%
$\mathcal{A}_{3}$&%
290.4\mbox{${}^{\scriptscriptstyle(e)}$}&210.1\mbox{${}^{\scriptscriptstyle(e)}$}&162.3\mbox{${}^{\scriptscriptstyle(e)}$}&122.5\mbox{${}^{\scriptscriptstyle(c)}$}&1.27\mbox{${}^{\scriptscriptstyle(e)}$}&1.92\mbox{${}^{\scriptscriptstyle(e)}$}&2.15\mbox{${}^{\scriptscriptstyle(e)}$}&0.99\mbox{${}^{\scriptscriptstyle(c)}$}\\
&$\mathcal{A}_{4}$&%
386.5\mbox{${}^{\scriptscriptstyle(c)}$}&350.3\mbox{${}^{\scriptscriptstyle(c)}$}&333.0\mbox{${}^{\scriptscriptstyle(c)}$}&136.7\mbox{${}^{\scriptscriptstyle(c)}$}&1.61\mbox{${}^{\scriptscriptstyle(c)}$}&1.49\mbox{${}^{\scriptscriptstyle(c)}$}&1.39\mbox{${}^{\scriptscriptstyle(c)}$}&2.09\mbox{${}^{\scriptscriptstyle(c)}$}\\
&$\mathcal{A}_{5}$&%
712.7\mbox{${}^{\scriptscriptstyle(g)}$}&739.6\mbox{${}^{\scriptscriptstyle(g)}$}&1112.9\mbox{${}^{\scriptscriptstyle(f)}$}&322.6\mbox{${}^{\scriptscriptstyle(c)}$}&1.09\mbox{${}^{\scriptscriptstyle(g)}$}&0.64\mbox{${}^{\scriptscriptstyle(g)}$}&0.63\mbox{${}^{\scriptscriptstyle(f)}$}&1.35\mbox{${}^{\scriptscriptstyle(c)}$}\\
&$\mathcal{A}_{6}$&%
1096.3\mbox{${}^{\scriptscriptstyle(f)}$}&1261.6\mbox{${}^{\scriptscriptstyle(d)}$}&1275.0\mbox{${}^{\scriptscriptstyle(d)}$}&1109.5\mbox{${}^{\scriptscriptstyle(f)}$}&0.78\mbox{${}^{\scriptscriptstyle(f)}$}&0.70\mbox{${}^{\scriptscriptstyle(d)}$}&0.66\mbox{${}^{\scriptscriptstyle(d)}$}&0.78\mbox{${}^{\scriptscriptstyle(f)}$}\\
\midrule\midrule
\end{tabular}
\end{table*}
%-------------------------------------------------------------------------------

\section{\label{sec:Conclusion}Conclusion}

All features of the experimental emission spectra of oligothiophenes with up
to five rings (i.e., up to $105$ vibrational DOFs) are well reproduced by
our OTF-AI-TGA calculations. The efficiency of the TGA formulation is found
to allow treating all vibrational DOFs on an equal footing even in case of
larger systems especially since the OTF-AI scheme does not require an
a~priori knowledge of the potential energy surfaces and the TGA approach
remains robust for floppy molecules. No symmetry considerations are
necessary; in particular, neither the dynamics nor the analysis relies on
any symmetry assumptions. Moreover, further considerable gain in efficiency
without loosing any substantial information can be obtained by employing
Hessian interpolation.

Experimentalists try, often successfully, to translate the spectral features
into a dynamical picture, which for theoreticians is often the starting
point. The extraction of the essential information from the dynamical
simulation, however, is often as difficult as the simulation itself. We
presented, therefore, a novel systematic approach to identify groups of
vibrations that are essential for the dynamics and for the spectrum. This
approach even allowed us to compare different oligothiophenes T$n$ and to
study changes in their spectra with increasing $n$:~Their vibrational line
shapes are modulated by inter-ring stretch and ring-squeeze vibrations, the
latter contributing to the spectral broadening for longer chains. The ground
and excited potential energy surfaces become more similar as the chain
length increases; this, in turn, reduces the amplitude of the dynamics
induced by emission and results in a shift of the intensity toward the $0$--$%
0$ transition. The phenomenon is also reflected in the different dependences
of the $0$--$0$ and vertical transition energies on $1/n$.

The OTF-AI-TGA scheme also allowed us, by evaluating the bond length
alternation, to study directly dynamical oscillations between the quinoid
and aromatic characters of individual rings in the oligothiophene chain.

OTF-AI-TGA is also useful as a preliminary test. The expensive OTF-AI
information stored during the TGA simulation can be reused in other
semiclassical methods such as poor person's Herman-Kluk (HK) propagator,
where the HK prefactor is for all contributing trajectories assumed to be
equal to the prefactor of the central trajectory.\cite{Tatchen:2011} In
systems, which are too large to be treated with a more sophisticated quantum
or semiclassical method, but for which the TGA is insufficient, e.g., due to
the importance of interference effects, the analysis of the OTF-AI-TGA
results can be used to define a subspace of reduced dimensionality, in which
the most important dynamics occurs. Within this subspace, the effects that
cannot be described with the TGA may be studied with less efficient yet
better-suited methods.\cite{Grossmann:2006} Alternative approaches for
constructing the information-flow matrix in order to maximize the decoupling
of the DOFs with minimal impact on the resulting spectrum are the subject of
our ongoing research.

Finally, let us note that the computational protocol presented here is not
limited to linear spectroscopy; non-linear spectra such as time-resolved
stimulated emission can also be evaluated with the OTF-AI-TGA. 
%-------------------------------------------------------------------------------

\begin{acknowledgments}
This research was supported by the Swiss National Science Foundation with
Grants No.~200020\_150098 and National Center of Competence in Research
(NCCR) Molecular Ultrafast Science \& Technology (MUST), and by the EPFL.
\end{acknowledgments}

%-------------------------------------------------------------------------------
\appendix
%-------------------------------------------------------------------------------

\section{\label{sec:rd}TGA in subspaces of reduced dimensionality}

%-------------------------------------------------------------------------------
One of the main goals of the normal mode analysis elaborated in Subsec.~\ref%
{subsec:nn_analysis_theory} is identifying the normal modes essential for
the dynamics. Restriction to these most important modes allows one to devise
a simplified model of reduced dimensionality, e.g., in the spirit of the
well-studied pyrazine vibronic coupling model.\cite{Stock_Woywod:1995}
Moreover, this reduction also broadens the class of computationally
available methods. After the reduction, one may be able to employ, e.g., the
Gaussian basis methods,\cite%
{Shalashilin_JCP_00,*Worth_Burghardt:2004,*Martinez_ACP_02} or various
approaches from the family of the semiclassical initial value representation.%
\cite{Miller:2001,*Herman:1994,*Thoss_Wang:2004,*Kay:2005}

Let us consider a system with $D$ vibrational DOFs. In a typical OTF-AI-TGA
calculation, one evolves the $D$-dimensional GWP by classically propagating
its center $x^t$ and by evaluating the phase factor $\gamma^{t}$ and the
complex time-dependent width matrix $A^{t}$ by means of Lee and Heller's $P$-%
$Z$ algorithm\cite{Lee_Heller:1982} summarized in Subsec.~\ref{subsec:TGA}
[Eqs.~(\ref{eq:TGA_gamma}) and (\ref{eq:TGA_APZ})].

As in Subsec.~\ref{subsec:nn_analysis_theory}, we identify the $D$%
-dimensional space of vibrational DOFs with the set $\mathcal{D}%
=\{1,\ldots,D\}$. We would like to take advantage of the stored $D$%
-dimensional trajectory information, and, at the same time, to restrict
ourselves to a subset $\mathcal{P}\subseteq\mathcal{D}$ of only $d<D$ most
important vibrational degrees of freedom and define a linear projection $%
\pi\!:\,\mathbb{R}^{D}\rightarrow\mathbb{R}^{d}$ from the full space of $D$
vibrational DOFs to the subspace of physical interest. Formally 
\begin{equation}  \label{eq:mapping_pi}
\pi_{ij}:=\delta_{\mathcal{P}_{i},j}\text{ for }1\leq{}i\leq{}d\text{ and }%
1\leq j\leq D,
\end{equation}
where $\mathcal{P}_{i}$ denotes the $i$th element of the ordered set $%
\mathcal{P}$.

The ``reduced'' $d$-dimensional GWP is again propagated using the $P$-$Z$
formalism. However, if $(q^{t},\,p^{t})$ denotes the trajectory followed by
the original, $D$-dimensional GWP, then the center of the reduced Gaussian
follows a classical trajectory $\left(\pi\cdot q^{t},\pi\cdot p^{t}\right)$ 
%$(I_{2}\otimes\pi)\cdot(q^{t},\,p^{t})$
in the reduced, $2d$-dimensional phase space. Also, the initial conditions
for the time-dependent $P^{t},\,Z^{t}$ matrices must be replaced with 
\begin{subequations}
\begin{align}
\bar{Z}^{0}&=\pi\cdot\pi^{\mathsf{T}}=I_{d}, \\
\bar{P}^{0}&=2i\hbar\,\pi\cdot A^{0}\cdot\pi^{\mathsf{T}}.
\end{align}
Here, $d$-dimensional matrices are denoted with a bar and $A^{0}$ is the
initial width matrix of the $D$-dimensional GWP.

Finally, we need to isolate the $\mathcal{P}$-contribution to the effective
Lagrangian $\mathcal{L}_{\text{eff}}$, which is required in Eq.~(\ref%
{eq:TGA_gamma}) for evaluating the time-dependent complex phase $\gamma^{t}$%
. This is conveniently done using conservation of energy, since in all our
calculations we consider only stationary initial states. Therefore 
\end{subequations}
\begin{equation}  \label{eq:cons_energy}
\frac{1}{2}{\zeta^t}^{\mathsf{T}}\cdot\zeta^{t}+V(\eta^t)=V(\eta^0),
\end{equation}
with $\zeta$ denoting momentum conjugated to $\eta$; mass factors do not
explicitly appear since $\eta$ is already mass-scaled. Using Eq.~(\ref%
{eq:cons_energy}), the $\mathcal{L}_{\text{eff}}$-contribution to $\gamma^t$
in Eq.~(\ref{eq:TGA_gamma}) then reads 
\begin{align}
\int_{0}^{t}\mathcal{L}_{\text{eff}}^{\tau}\,d\tau &=\int_{0}^{t}\!\bigl[%
\frac{1}{2}({\zeta^{\tau}})^{\mathsf{T}}\cdot\zeta^{\tau}-V(\eta^{\tau})%
\bigr]d\tau  \notag \\
&=\int_{0}^{t}\!\bigl[({\zeta^{\tau}})^{\mathsf{T}}\cdot\zeta^{\tau}-V(%
\eta^0)\bigr]d\tau  \notag \\
&=\int_{0}^{t}\!({\zeta^{\tau}})^{\mathsf{T}}\cdot\zeta^{\tau}\,d\tau-V(%
\eta^0)\,t.
\end{align}
The part of this expression pertinent to the dynamics within the subset of
vibrational DOFs $\mathcal{P}$ is then easily obtained by replacing $%
\zeta^{t}$ with $\pi\cdot\zeta^{t}$, i.e., 
\begin{equation}
\left.\int_{0}^{t}\mathcal{L}_{\text{eff}}^{\tau}\,d\tau\right|_{\mathcal{P}%
} =\int_{0}^{t}\!(\pi\cdot{\zeta^{\tau}})^{\mathsf{T}}\cdot(\pi\cdot\zeta^{%
\tau})\,d\tau-V(\eta^0)\,t.
\end{equation}
The term $V(\eta^0)\,t$ generates an overall phase depending linearly on $t$
and is responsible only for shift of the resulting spectrum without altering
its shape. 
%-------------------------------------------------------------------------------

\section{\label{sec:ecc}Analysis of the effective conjugation coordinate}

%-------------------------------------------------------------------------------
Oligomer spectra are usually analyzed in terms of the so-called \textit{%
effective conjugation coordinate}\cite%
{Zerbi:1989,Geisselbrecht:1993,Zerbi:2007} (ECC), i.e., the totally
symmetric internal coordinate the excitation of which triggers the
conformational change between the aromatic to the quinoid structures of the
molecule. This approach is especially popular within Raman spectroscopy.\cite%
{Navarette:1991b,Zerbi_Inganas:1991,Castiglioni:2004,Orti:2005,Fazzi:2010}
For the oligothiophene family T$n$, ECC captures the alternation between
adjacent bonds and is defined as 
\begin{equation}  \label{eq:R}
\mbox{\usefont{T2A}{\rmdefault}{m}{n}\CYRYA}:=\frac{1}{\sqrt{\bar{N}}}%
\sum_{a=1}^{\bar{N}}(-1)^{a-1}r_{a},
\end{equation}
where $r_{a}$ is the Cartesian vector connecting the $a$th and $(a+1)$th
carbon atoms of the backbone comprised of $\bar{N}=4n-1$ C-C bonds in total.
Further insight is gained by restating Eq.~(\ref{eq:R}) in the normal-mode
coordinates. By employing transformation (\ref{eq:Cartesian_to_normalModes}%
), we obtain 
\begin{equation}
\mbox{\usefont{T2A}{\rmdefault}{m}{n}\CYRYA}=\mbox{\usefont{T2A}{%
\rmdefault}{m}{n}\CYRYA}_{\text{ref}}+\mathrm{R}\cdot\eta\text{, with }%
\mathrm{R}:=\mathrm{S}\cdot T\text{ and }\mbox{\usefont{T2A}{%
\rmdefault}{m}{n}\CYRYA}_{\text{ref}}:=\mathrm{S}\cdot\xi_{\text{ref}},
\end{equation}
where $T$ is the transformation matrix of Eq.~(\ref%
{eq:Cartesian_to_normalModes}), $\xi_{\text{ref}}$ denotes Cartesian
coordinates of a reference geometry, and 
\begin{equation}
\mathrm{S}:=\frac{1}{\sqrt{\bar{N}}}\sum_{a=1}^{\bar{N}}(-1)^{a}(\mathrm{P}%
_{a}-\mathrm{P}_{a+1})
\end{equation}
is a generalization of the projector $\mathrm{P}_{b}$ defined below Eq.~(\ref%
{eq:Eckart_conditions}).

Then, the normalized \textquotedblleft coupling strength\textquotedblright\ $%
\nu ^{j}$ of the $j$th normal mode to $\mbox{\usefont{T2A}{\rmdefault}{m}{n}%
\CYRYA}$ reads 
\begin{equation}
\nu ^{j}:=\frac{\mathop{\mathrm{Tr}}\mathrm{Y}_{j}^{\mathsf{T}}\!\cdot 
\mathrm{R}^{\mathsf{T}}\!\cdot \mathrm{R}\cdot \mathrm{Y}_{j}}{%
\mathop{\mathrm{Tr}}\mathrm{R}^{\mathsf{T}}\!\cdot \mathrm{R}},
\label{eq:app:nu}
\end{equation}%
where the square matrix $\mathrm{Y}_{j}$ is defined as $(\mathrm{Y}%
_{j})_{kl}:=\delta _{jk}\delta _{jl}$.

However, the quantity $\mbox{\usefont{T2A}{\rmdefault}{m}{n}\CYRYA}$ changes
during the dynamics and its variations are described in terms of 
\begin{equation}
\delta\mbox{\usefont{T2A}{\rmdefault}{m}{n}\CYRYA}^{t}:=\mbox{\usefont{T2A}{%
\rmdefault}{m}{n}\CYRYA}^{t}-\mbox{\usefont{T2A}{\rmdefault}{m}{n}\CYRYA}%
^{0}=\mathrm{R}\cdot{\bar{\eta}^{t}}\text{, with }\bar{\eta}%
^{t}:=(\eta^{t}-\eta^{0}).
\end{equation}
Now, in order to asses the importance of a particular normal mode with
respect to $\delta\mbox{\usefont{T2A}{\rmdefault}{m}{n}\CYRYA}^{t}$, we can
not use Eq.~(\ref{eq:app:nu}) directly, since $\nu^{j}$ provides only a
static picture. To remedy this, we introduce a more appropriate measure of
dynamical coupling: 
\begin{equation}  \label{eq:app:upsilon}
\upsilon^{j}:=\frac{\varsigma^{j}}{\sum_{k}\varsigma^{k}} \text{, with }
\varsigma^j:=\int_{0}^{t}\!dt\,\bar{\eta}^{t}\cdot\mathrm{Y}_{j}^\mathsf{T}%
\!\cdot\mathrm{R}^\mathsf{T}\!\cdot\mathrm{R}\cdot\mathrm{Y}_{j}\cdot\bar{%
\eta}^{t},
\end{equation}
where the summation runs over all normal modes.

\begin{figure}
\centering
\includegraphics{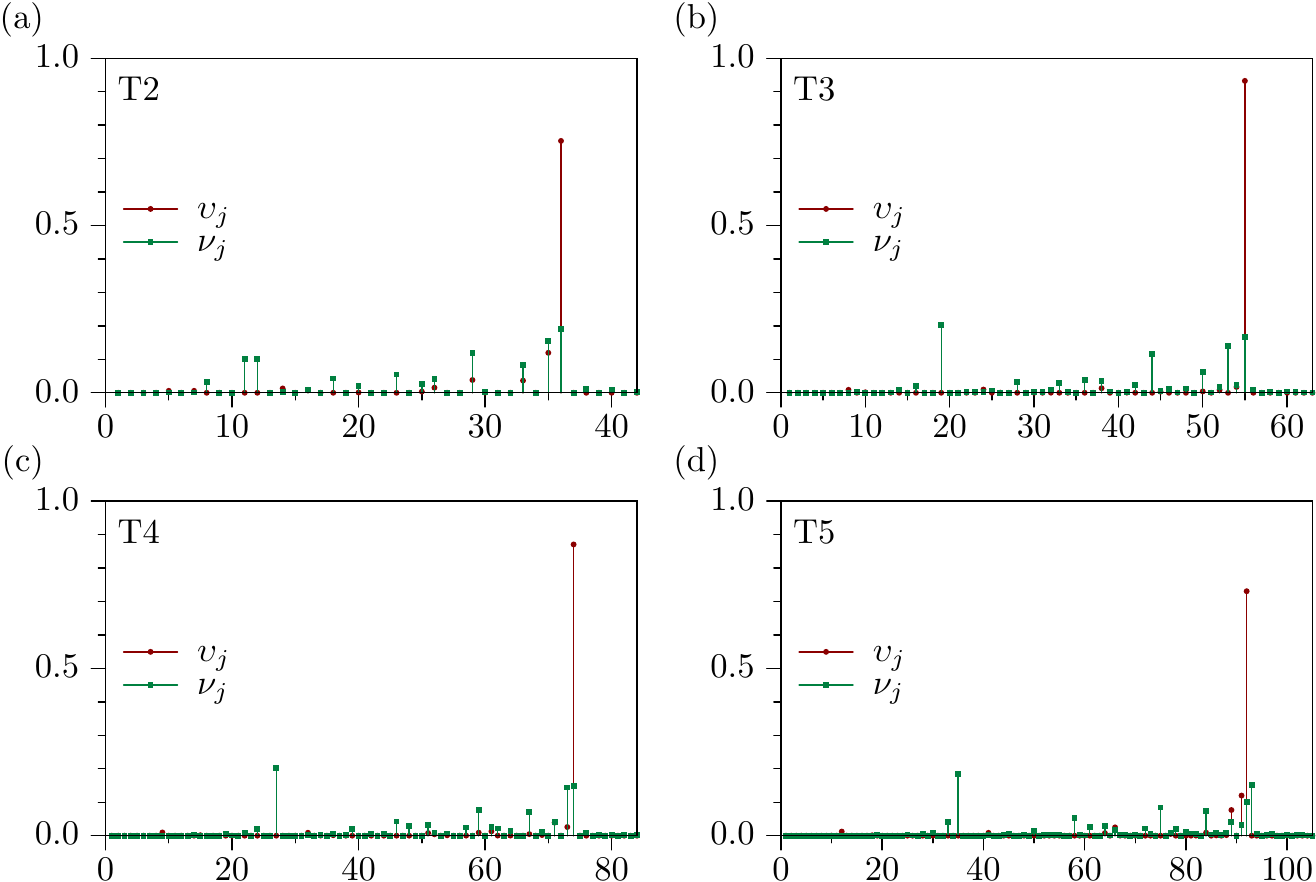} 
\caption{\label{fig:ecc}
Comparison of the static, $\nu_{j}$ [see Eq.~(\ref{eq:app:nu})], and dynamic, $\upsilon_{j}$ [see Eq.~(\ref{eq:app:upsilon})],
contribution of individual normal modes to the ECC $\RR$ of Eq.~(\ref{eq:R}) for the oligothiophenes T$n$, $n\in\{2,3,4,5\}$.
}
\end{figure}

A~comparison of individual normal modes for T$n$, $n\in\{2,3,4,5\}$, in
terms of $\nu^{j}$ and $\upsilon^{j}$ is shown in Fig.~\ref{fig:ecc}, which
demonstrates clearly that only certain modes contributing to $%
\mbox{\usefont{T2A}{\rmdefault}{m}{n}\CYRYA}$ are excited during the
fluorescence process. This means that an analysis based merely on $\nu^{j}$
would be incomplete.

In Subsec.~\ref{subsec:nn_analysis_theory}, individual normal modes were
classified into independent groups $\mathcal{A}_{i}$ [see Eq.~(\ref%
{eq:active_groups})]. Using Eq.~(\ref{eq:app:upsilon}), we can estimate the
dynamical influence of a particular group $\mathcal{A}_{i}$ on $\delta%
\mbox{\usefont{T2A}{\rmdefault}{m}{n}\CYRYA}^{t}$ by employing 
\begin{equation}  \label{eq:kappa}
\kappa_{i}:=\sum_{j\in\mathcal{A}_{i}}\upsilon^{j}.
\end{equation}
Table \ref{tab:ecc_frac} demonstrates that variations in $%
\mbox{\usefont{T2A}{\rmdefault}{m}{n}\CYRYA}$ can be assigned mostly to the
group $\mathcal{A}_{1}$, and, hence, the $\alpha$-peaks (see Fig.~\ref%
{fig:spec}) originate from the change of the ECC during the dynamics induced
by the fluorescence process.

%-------------------------------------------------------------------------------

\begin{table}[h!t]
\begin{ruledtabular}
\begin{tabular}{ccccccc}
& $\mathcal{A}_{1}$ & $\mathcal{A}_{2}$ & $\mathcal{A}_{3}$ &$\mathcal{A}_{4}$ & $\mathcal{A}_{5}$ & $\mathcal{A}_{6}$\\\hline
T$2$ & 0.947 & 0.013 & 0.006 & 0.006 & 0.008 & 0.015 \\
T$3$ & 0.962 & 0.010 & 0.009 & 0.001 & 0.000 & 0.003 \\
T$4$ & 0.942 & 0.009 & 0.010 & 0.002 & 0.004 & 0.013 \\
T$5$ & 0.927 & 0.008 & 0.000 & 0.012 & 0.001 & 0.025 \\
\end{tabular}
\end{ruledtabular}
\caption{\label{tab:ecc_frac}
Contribution of the $i$th group $\mathcal{A}_{i}$ to ECC in terms of $\kappa_{i}$ introduced in Eq.~(\ref{eq:kappa}).
}
\end{table}

%-------------------------------------------------------------------------------
\bibliographystyle{aipnum4-1}

\end{document}